\documentclass[letterpaper,twocolumn,10pt]{article}
\usepackage{usenix}
\usepackage{cite}
\usepackage{tikz}
\usepackage{amsmath}
\usepackage{amsthm}
\usepackage{amssymb}
\usepackage{multirow}
\usepackage{tabularx}
\usepackage{booktabs}
\usepackage{multirow} 
\usepackage{multicol}
\usepackage{fancyhdr}
\usepackage{comment}
\usepackage{microtype}
\usepackage{threeparttable}
\usepackage{graphicx}
\usepackage{listings}
\usepackage{subfigure}
\usepackage{url}
\usepackage{tcolorbox}
\usepackage{pifont}

\usepackage{algorithm}
\usepackage[noend]{algpseudocode}
\usepackage{usenixbadges}
\renewcommand{\arraystretch}{1.3}

\algnewcommand{\Indent}[1]{\State\hspace{#1}}

\newcommand{\revise}[1]{{\color{black}#1}}

\newenvironment{packeditemize}{
\begin{list}{$\bullet$}{
\setlength{\labelwidth}{6pt}
\setlength{\itemsep}{2pt}
\setlength{\leftmargin}{\labelwidth}
\addtolength{\leftmargin}{\labelsep}
\setlength{\parindent}{1pt}
\setlength{\listparindent}{\parindent}
\setlength{\parsep}{1pt}
\setlength{\topsep}{1pt}}}{\end{list}}

\usepackage{makecell,multirow}
\usepackage{enumitem}

\usepackage{tabularx}
\usepackage{array}

\usepackage{booktabs}  
\usepackage{xcolor}    
\usepackage{makecell}  
\usepackage{caption}
\usepackage{needspace}
\usepackage{circledsteps}
\usepackage[normalem]{ulem}

\newcommand{\BAdown}[3]{#1\,{$\rightarrow$}\,#2\textsuperscript{\textcolor{red}{\scriptsize$\downarrow$\,#3}}}
\newcommand{\BAup}[3]{#1\,{$\rightarrow$}\,#2\textsuperscript{\textcolor{green!60!black}{\scriptsize$\uparrow$\,#3}}}

\newcommand{\TN}{BenchChecker}

\lstdefinelanguage{arxivdiff}{
  morecomment=[f][\color{green!40!black}]{+},
  morecomment=[f][\color{red!70!black}]{-},
  morecomment=[f][\color{blue!65!black}]{@@}
}
\lstdefinestyle{arxivdiff}{
  language=arxivdiff,
  basicstyle=\scriptsize\ttfamily,
  columns=fullflexible,
  keepspaces=true,
  showstringspaces=false,
  breaklines=true,
  breakatwhitespace=false,
  numbers=none,
  frame=lines,
  tabsize=4,
  xleftmargin=0pt
}

\begin{document}
\pagestyle{empty}

\begin{textblock}{7.0}(0.75,0.3)
  \vbox to 0pt{
    \vss
    \noindent
    \makebox[7in][r]{
      \raisebox{-\height}{\includegraphics[width=0.6in]{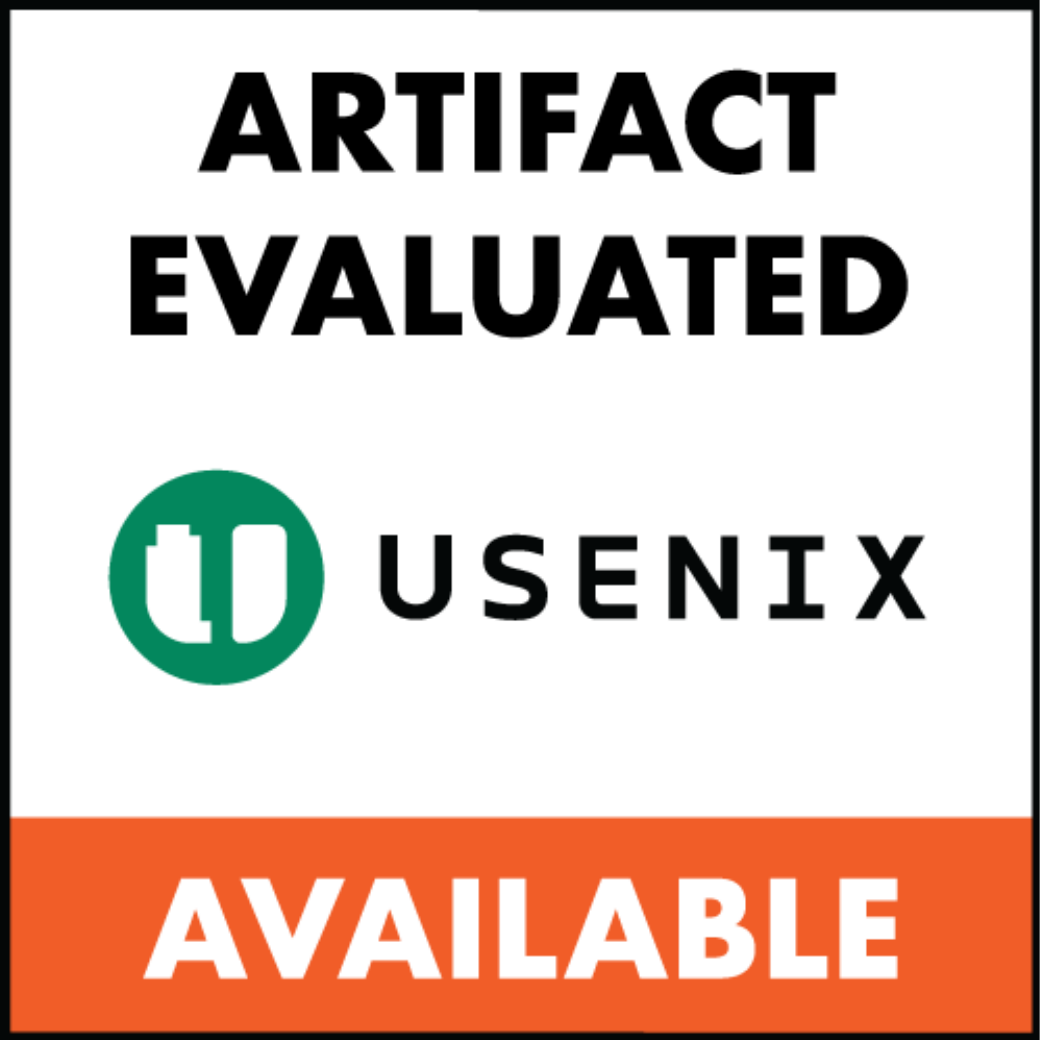}}
      \hspace{10pt}
      \raisebox{-\height}{\includegraphics[width=0.6in]{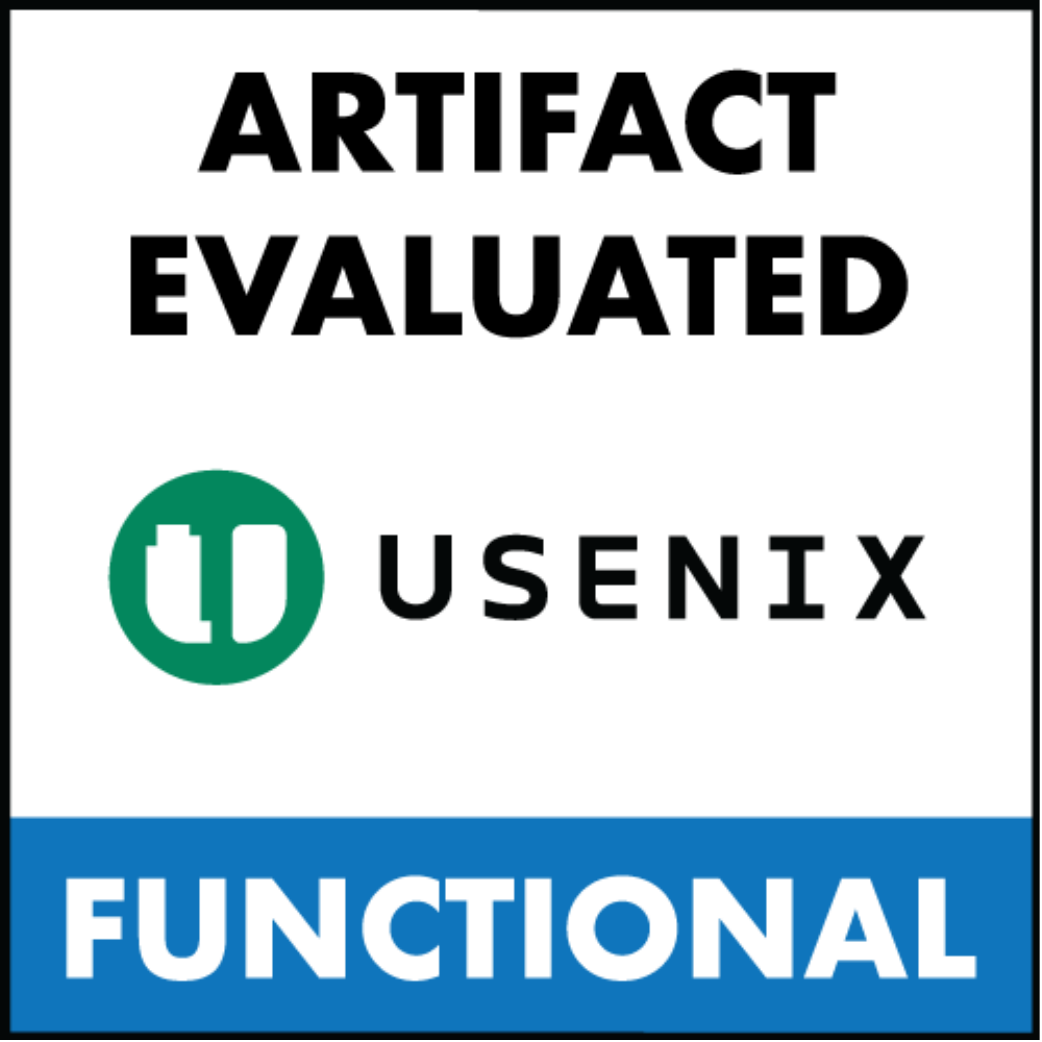}}
    }
    \vskip-\prevdepth}
\end{textblock}

\title{\TN{}: Assessing the Credibility of Bug-Fixing Benchmarks for LLMs}

\author{
Di Wu\textsuperscript{1}, Xu He\textsuperscript{1}, Shu Wang\textsuperscript{2}, and Kun Sun\textsuperscript{1}\\
\textsuperscript{1}George Mason University, \textsuperscript{2}Palo Alto Networks, Inc.
}

\maketitle
\thispagestyle{empty}

\begin{abstract}

While Large Language Models (LLMs) show promise for Automatic Program Repair (APR), existing benchmarks suffer from \revise{data contamination, in which} test samples appear in training sets, artificially inflating performance. To address this, we propose \TN{}, a framework compatible with both open-source and closed-source models that supports more credible evaluation by detecting such contamination. Grounded in the observation that persistent patches likely remain in recent repository snapshots used for training, BenchChecker employs a forensics-based two-stage detection process using only textual model outputs together with public repository history. The first stage conducts a repository presence test by probing the model with code prefixes to infer training inclusion; the second stage performs a patch presence test to verify patch persistence via similarity alignment and test-suite verification. Evaluations using StarCoder-15B, StarCoder2-15B, and CodeGen2-16B trained on The Stack v2 demonstrate that \TN{} achieves robust detection with AUC scores around 0.9 across six benchmarks. Furthermore, applying \TN{} to SWE-bench Verified and Multi-SWE-bench reveals that filtering contaminated samples reduces the reported resolution rates of most LLMs by more than 20\% relative to the original rates on medium-difficulty tasks.

\end{abstract}

\section{Introduction}

Large language models (LLMs) have demonstrated strong capabilities in software-related tasks\revise{~\cite{zhang2026chatgpt}}, including automated bug fixing in repositories~\cite{chen2024large}. Consequently, bug-fixing benchmarks~\cite{jimenez2023swe, just2014defects4j, zan2025multi, widyasari2020bugsinpy} have emerged to systematically evaluate recent advances of LLMs in bug fixing~\cite{swebench_leaderboards}. However, the issue of data contamination~\cite{magar2022data} raises concerns about the credibility of benchmark results~\cite{magar2022data}. When benchmark patches are present in a model’s training dataset, the model may rely on memorization rather than reasoning or generalization to repair the samples. This issue has already affected practice: OpenAI stopped reporting SWE-bench Verified scores due to contamination concerns \cite{openai_swebench_verified_2026}. 

Data contamination in bug-fixing benchmarks arises from two main factors. First, widely used bug-fixing benchmarks are typically derived from popular open-source repositories, which are frequently incorporated into large-scale training datasets. Our analysis reveals that for six commonly used benchmarks, over 85\% of their source repositories appear in The Stack v2~\cite{lozhkov2024starcoder}, \revise{the largest publicly available code training dataset to date} (see Section~\ref{sec:datasets}). Second, when such repositories are included in the training dataset, the corresponding patches often remain in those snapshots. We further verify that for these six benchmarks, more than 50\% of the test samples’ patches remain persistent in the corresponding repositories (see Table~\ref{tab:bugfix-stats}).

There are two unique challenges in assessing \revise{dataset contamination in bug-fixing benchmarks}. First, LLMs typically do not disclose their training datasets, preventing direct verification of whether the patches used in evaluation benchmarks were included in the models’ training \revise{sets}~\cite{meta2025llama4}. Second, closed-source LLMs~\cite{achiam2023gpt, anthropic2023claude} offer access only to textual outputs rather than intermediate representations, such as logits. Consequently, existing membership-inference (MI) based detection methods~\cite{carlini2021extracting, zhang2024min}, which depend on confidence scores derived from logits to determine whether a sample is part of the training set, cannot be directly applied.

To address these challenges, we introduce \TN{}, a \textit{forensics-based} two-stage framework designed to detect contaminated test samples and assess their impact on bug-fixing benchmarks for LLMs. \TN{} stems from the observation that \revise{code LLM training sets typically include} recent repository snapshots, not only historical ones, to maintain data freshness. For a sample patch that was introduced early and was never removed before the cutoff date, the patch is highly likely to be contained in these snapshots, resulting in contamination. Thus, the first stage is \revise{a} \textit{repository presence test} to determine if the sample's repository appears in the model’s training set, while the second stage is \revise{a} \textit{patch presence test} to examine whether the sample's patch remains persistent in the repository before the training cutoff date.

In the first stage, we formulate the problem as a Bernoulli hypothesis test by partitioning the repository into prefix queries and suffix targets, defining a hit event as the exact reproduction of the target. Given that repositories seen during training yield higher hit probabilities, multiple hits serve as strong evidence of inclusion. To ensure validity, we select code snippets based on three criteria: \textit{repetitiveness}, ensuring the snippet occurs frequently to facilitate memorization; \textit{distinctiveness}, requiring the target to be rare in a large reference corpus to rule out general knowledge; and \textit{context-independence}, requiring high conditional entropy to minimize contextual predictability. Specifically, this stage consists of three steps. First, we identify repetitive syntactic patterns to derive queries. Second, we filter them via a reference model based on entropy. Third, we probe the target LLM to compute a $p$-value reflecting the likelihood of training data inclusion.

Once the inclusion of the repository is confirmed, the second stage assesses whether early patches persist in the repository snapshot at the training cutoff date (cutoff snapshot). First, we pre-process the patch code by formatting and removing non-code content. Second, we identify the correspondence of each patch block to the cutoff snapshot. Third, we compute code similarity for each matched pair of blocks. If all patch blocks have highly similar counterparts, the patch is considered to be present. Fourth, when similarity matching is inconclusive, we test whether the corresponding target-snapshot blocks can successfully pass the sample's test suite. Passing all tests indicates that the cutoff snapshot contains an equivalent implementation to the patch.

To evaluate the effectiveness of \TN{} in controlled settings, we test it on three open-source LLMs: \textit{StarCoder-15B}, \textit{StarCoder2-15B}, and \textit{CodeGen2-16B}, all of which are trained on \textit{The Stack v2} dataset, where the ground-truth sample membership is known. Across six bug-fixing benchmarks, we compare \TN{} against four logit-based baselines. The results show that \TN{} consistently achieves an AUC of around 0.9 with a False Positive Rate below 0.1, significantly outperforming existing methods. These findings confirm that \TN{} effectively detects contaminated samples using only textual outputs, rendering it suitable for both open-source and closed-source models.

We apply \TN{} to assess the reliability of SWE-bench Verified~\cite{jimenez2023swe} and Multi-SWE-bench~\cite{zan2025multi} using their leaderboard submission results~\cite{swebench_leaderboards, multiswebench_leaderboard}, where \TN{} remains effective while logit-dependent methods fail due to \revise{the restricted output access of} SOTA closed-source LLMs. After filtering contaminated samples, we observe that resolution rates for the top-10 submissions decrease by more than 20\% relative to their original rates on medium-difficulty tasks (specifically the \texttt{1-4 h} subset in SWE-bench Verified and the \texttt{Medium} subset in Multi-SWE-bench), showing substantial overfitting in current evaluations. Furthermore, this contamination-induced degradation declines from 40\% in 2023 to 5\% in 2025, suggesting that newer LLMs demonstrate substantially stronger generalization capabilities.

In summary, our paper makes the following contributions:

\begin{packeditemize} \item We introduce the first framework for detecting data contamination in LLM bug-fixing benchmarks. Distinct from prior approaches, our method relies solely on textual outputs to support both open-source and closed-source models.

\item Through comprehensive evaluations, we demonstrate that our two-stage framework significantly outperforms existing logit-dependent baselines in identifying contaminated samples across multiple benchmarks.

\item Our empirical analysis on two widely used benchmarks reveals that the resolution rates reported on leaderboards are significantly overestimated. \end{packeditemize}

\section{Background}
\label{sec:background}

\noindent {\bf Bug-Fixing Benchmark.}
A bug-fixing benchmark is a curated dataset and evaluation protocol\revise{~\cite{afrose2023evaluation}} for measuring how effectively a model repairs software defects. Such a dataset typically includes: (1) real-world bugs drawn from actual source-code repositories, including the buggy version and the developer-provided patch; (2) a test suite configured so that some tests fail on the buggy version (fail-to-pass tests) and some continue to pass after the fix (pass-to-pass tests), validating both the repair and regression avoidance; and (3) metadata such as the programming language, repository identity, issue description, and commit details.
Popular benchmarks such as SWE-bench~\cite{jimenez2023swe} and Multi-SWE-bench~\cite{zan2025multi} maintain leaderboards that track and compare the resolution rates of state-of-the-art (SOTA) LLMs~\cite{swebench_leaderboards, multiswebench_leaderboard}.

\noindent {\bf \revise{Code LLM Training Sets.}} To maintain data freshness and mitigate temporal distribution shifts~\cite{lazaridou2021mind}, pretraining datasets for both open-source and closed-source Code LLMs incorporate recent repository snapshots up to the training cutoff date~\cite{agarwal2022temporal}. \revise{This follows common practice in LLM pretraining, where relying on outdated data can degrade performance~\cite{lazaridou2021mind}, especially by causing code LLMs to hallucinate deprecated APIs rather than use the latest versions~\cite{liang2025rustevo}.} For example, mainstream open-source datasets such as The Stack~\cite{li2023starcoder} and The Stack v2~\cite{lozhkov2024starcoder} follow this principle by collecting the latest repository snapshots from default branches. However, this practice inadvertently introduces data contamination risks for bug-fixing benchmarks, as these snapshots already contain the ground-truth patches of bug samples.

\section{System Design}
\label{sec:system design}

\subsection{Motivation}

Since closed-source LLMs do not disclose their training sets and only offer textual outputs~\cite{anthropic2023claude, gpt5}, we formulate the problem as follows: given a bug-fixing sample (comprising a patch and its source repository) and a target LLM with a known training cutoff date but an unknown training set, our goal is to determine whether the patch appears in the LLM's training data. The system operates under a black-box setting, where only the model's textual outputs are accessible. 

\revise{
A straightforward approach is to query the model with bug samples and check whether the generated patches match the ground-truth patches. However, it is unreliable in practice. First, a model can infer a similar patch from contextual information (e.g., issue descriptions), yielding false positives; prior work reports that similarity alone cannot distinguish memorization from generalization~\cite{zhang2025memorize}. Second, a model might memorize the fixing logic yet output a distinct but equivalent fix, yielding false negatives. This problem is further pronounced in modern agentic repair systems, where long interaction trajectories drastically reshape the final diff~\cite{wang2025solvedissues,ceka2025traceability}. While logit-based membership inference offers a model-based alternative~\cite{carlini2021extracting,zhang2024min}, it remains inapplicable to black-box models that restrict access to textual outputs alone~\cite{achiam2023gpt,anthropic2023claude}.
}

Instead, we leverage the observation that dataset curators typically collect snapshots near the cutoff to maximize freshness, as described in Section~\ref{sec:background}. Consequently, patches merged prior to the cutoff are likely preserved in these recent snapshots, leading to contamination. We therefore focus on detecting contaminated samples satisfying two necessary conditions: (1) the repository is included in the training set, and (2) the target patch was merged and persisted before the cutoff. This approach decomposes the contamination detection problem into two tractable sub-problems: a \textit{repository presence test} to determine if the repository appears in the training set, and a \textit{patch presence test} to verify if the patch persists in the repository prior to the cutoff. \revise{Unlike the model-based methods discussed above, our \textit{forensics-based} approach combines multiple external information sources to build credible evidence for contamination detection rather than relying only on model information. Specifically, this approach assumes access to benchmark metadata, open-source code history, and model cutoff dates. These requirements are practical: benchmark metadata and open-source code histories are publicly available, and model cutoff dates are usually disclosed or can be estimated from release dates.} 

For the second sub-problem, rather than exhaustively tracking every historical commit, we adopt a more efficient strategy by directly verifying the patch's existence in the latest snapshot before the cutoff (cutoff snapshot). This simplification is supported by empirical evidence suggesting that the removal and subsequent re-introduction of patches is a rare edge case~\cite{romano2018multi}. Consequently, the presence of a patch in the cutoff snapshot serves as a strong indicator of its continuous persistence since its introduction.

\begin{figure}[t]
    \centering
    \includegraphics[width=\linewidth]{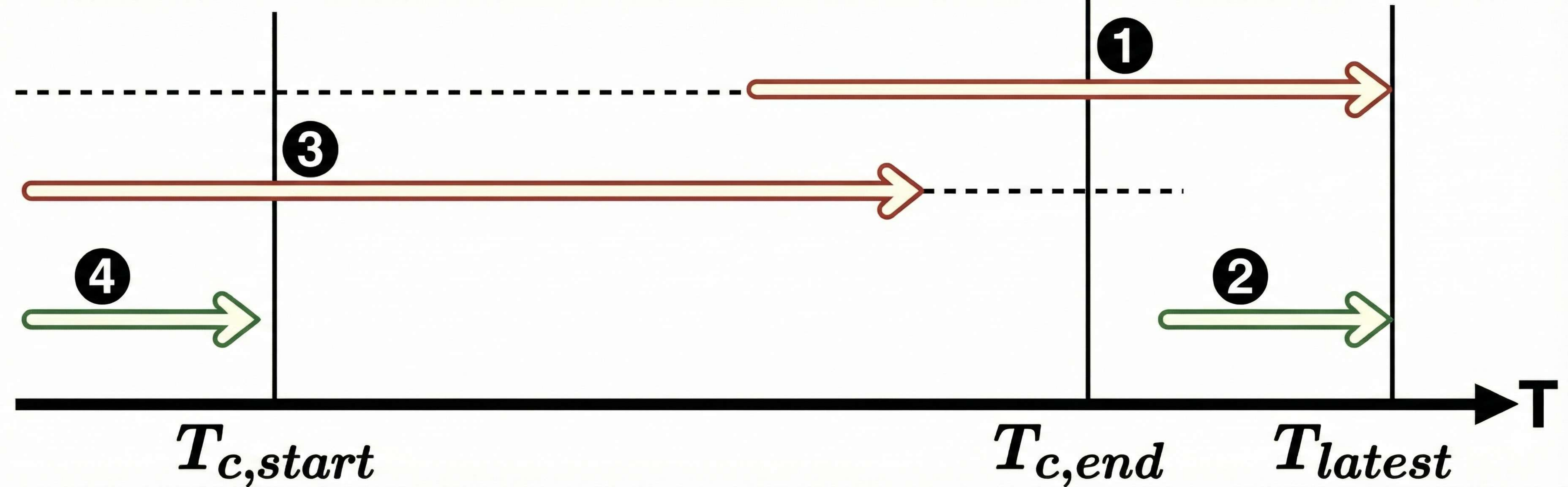} 
    \caption{Overlap between patch lifecycles and the \revise{training snapshot collection} interval.}
    \label{fig:timeline}
\end{figure}

To demonstrate the capability boundaries of our approach, we classify samples into four categories based on the temporal overlap between the patch lifecycle and the LLM snapshot collection interval, as illustrated in Figure~\ref{fig:timeline}. Specifically, let $[T_{c,start}, T_{c,end}]$ denote the collected snapshot interval, where $T_{latest}$ represents the timestamp of the cutoff snapshot. For a given patch, $T_{p,intro}$ marks its introduction, while $T_{p,persist}$ indicates the last timestamp of its existence before the cutoff. Accordingly, a sample is defined as ground-truth \textit{contaminated} if the patch lifecycle overlaps with the collection interval; otherwise, it is considered \textit{clean}.

\begin{packeditemize}
    \item[] \ding{172} $(T_{p,intro} \leq T_{c,end}) \land (T_{p,persist} \geq T_{latest})$: The patch is introduced before the collection ends and persists until the cutoff snapshot. In this case, the patched snapshots overlap with the collection interval, which will be correctly detected as a true positive contaminated sample.

    \item[] \ding{173} $(T_{p,intro} > T_{c,end}) \land (T_{p,persist} \geq T_{latest})$: The patch is introduced in the gap between the collection end and the cutoff snapshot. In this case, there is no overlap between the patched snapshots and the collection interval, but it will still be detected as a contaminated sample, resulting in a false positive. However, since dataset curators typically prioritize recent data, the interval between $T_{c,end}$ and $T_{latest}$ is usually minimal, causing its effect to be negligible. Thus, \revise{samples detected by our method} are highly probable to be actual contaminated samples.

    \item[] \ding{174} $(T_{p,persist} \geq T_{c,start}) \land (T_{p,persist} < T_{latest})$: The patch persists after the collection starts but is removed before the collection ends. In this case, there is indeed a contaminated sample that fails to be detected, resulting in a false negative. This is due to the practical unavailability of the exact $T_{c,start}$. Nevertheless, our empirical results indicate that such cases constitute only approximately 10\% of the samples. Furthermore, the significant degradation in model performance observed in the decontaminated subset supports the validity of our approach, despite these potential false negatives.

    \item[] \ding{175} $T_{p,persist} < T_{c,start}$: The patch is removed before the collection starts. In this case, the sample will be correctly detected as clean (i.e., a true negative). Our approach primarily aims to identify these samples to enable a trustworthy evaluation of models on the original benchmark without requiring additional post-cutoff data.
    
\end{packeditemize}

\begin{figure*}[t]
    \centering
    \includegraphics[width=\textwidth]{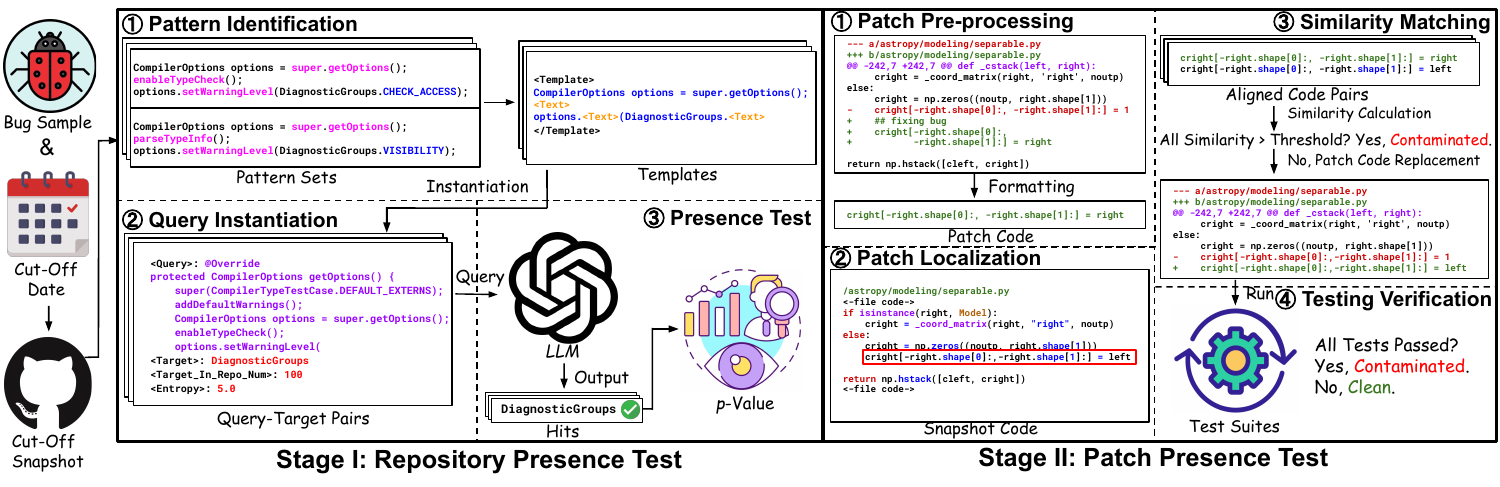}
    \caption{Workflow of \TN{}.}
    \label{fig:architecture}
\end{figure*}

\subsection{System Overview}

The overall workflow of \TN{} is illustrated in Figure~\ref{fig:architecture}. The first stage consists of three steps: {\em pattern identification}, {\em query instantiation}, and {\em presence test}. First, we identify repetitive patterns within the repository and group the corresponding code snippets into pattern sets, from which the common content is extracted as templates. Second, we instantiate the templates into target-query pairs, where the target is a specific word within the template, and the query is the prefix of the original code snippet ending immediately before the target. Third, we probe the model using all queries and record the number of successful target predictions (hits). A \textit{p-value} is computed from the hit count to quantify evidence that the repository appears in the model’s training set. This stage requires only a single repository snapshot. Here, we select the cutoff snapshot as the input for clarity.

Based on the repository inclusion results of the first stage, the second stage consists of four steps: \textit{patch preprocessing}, \textit{patch localization}, \textit{similarity matching}, and \textit{testing verification}. First, we clean and normalize the patch to extract its code blocks. Next, we align these patch blocks with the corresponding code regions in the cutoff snapshot to form block pairs. Similarity scores are then computed for each pair; if all scores exceed a predefined threshold, the sample is deemed contaminated. When similarity scores are inconclusive, we construct a modified diff by substituting patch blocks with their aligned snapshot counterparts, apply the diff to the buggy version, and run the sample’s test suite. If the modified diff passes all tests, the sample is classified as contaminated, indicating that an equivalent implementation exists in the cutoff snapshot.

\subsection{Stage I: Repository Presence Test}

\noindent {\bf Task Modeling.} Given only textual access to model $M$, we formulate repository-presence testing as follows: we treat repository $R$ as a collection of code snippets $C$. For each code snippet $C$ from $R$, we split it into a prefix query $q$ and a suffix target $t$, such that \( C = (q, t) \).
We query the model with $q$ and check whether it can accurately generate $t$. \revise{For each query, we generate a large response candidate set and assume that, when $M$ has memorized $t$, at least one candidate from this set can reproduce $t$ exactly, as reported by prior work~\cite{hayes2024measuring}.} Specifically, let \(\text{Resp}(M; q)\) denote the set of responses generated by model $M$ when queried with $q$. We define a {\em hit} as the event where \( t \) is included in \(\text{Resp}(M; q)\). Based on the observed hit events, we can infer whether repository $R$ is included in model $M$’s training set $D$. 

This task modeling is based on a key assumption: \textit{the probability of observing a hit event by $M$ is higher in a seen repository than in an unseen one}. This assumption is supported by empirical evidence that code-trained LLMs often memorize code verbatim~\cite{yang2024unveiling}, and the memorized content can be extracted using carefully crafted prompts~\cite{huang2407demystifying}. However, three specific scenarios may lead to false positives:

\begin{packeditemize}
    \item {\em Unmemorized Code Snippets:}  A single repository may contain thousands of code snippets, and the model is unlikely to memorize them equally, particularly when the snippet appears only once in the model’s training dataset. 
    \item {\em Prevalent Code \revise{Patterns}:} The target \( t \) may be a highly frequent code pattern commonly found across many repositories (e.g., pointer null checks), making it easily predictable and not tied to a specific repository. 
    \item {\em Strong Contextual Cues:} Real-world code often exhibits structural redundancy, leading to similar syntax and semantics across different code blocks, which in turn allows the model to infer \( t \) from the context of the query \( q \) rather than from the memorization of training data. 
    
\end{packeditemize}

\noindent {\bf Signature Snippets.} To mitigate these potential false positives, we introduce a class of effective code snippets called  \textit{signature snippets} that are selected based on three key criteria:

\begin{packeditemize}
    \item \textit{Repetitiveness:} The code snippet \( C \) must appear multiple times within the repository \( R \). Formally, \( C\) should satisfy:
    \begin{equation}
    \mathrm{count}_{R}(C) > 1,
    \end{equation}
    where $\mathrm{count}_{R}(C)$ denotes the number of times the code snippet \(C\) appears in \( R \). This follows prior findings that repeated data are more likely to be memorized by models~\cite{hayes2024measuring, biderman2023emergent, carlini2021extracting, yang2024unveiling}.

    \item \textit{Distinctiveness:} The target \( t \) must occur in a limited number of repositories, suggesting that it is not a commonly used code snippet. Let $\mathcal{R}$ denote the set of all reference repositories; the distinctiveness is defined as:
    \begin{equation}
    \sum_{r \in \mathcal{R}} \mathbf{1}[t \in r] \le \tau,
    \end{equation}
    where $\mathbb{I}[t \in r]$ is an indicator function equal to 1 if $t$ appears in repository $r$, and 0 otherwise. $\tau$ is a predefined threshold specifying the maximum allowable number of repositories in which $t$ can occur.
    
    \item \textit{Context-Independence}: The query \( q \) should provide no obvious cues about the target \( t \), thereby reducing the likelihood that the model's prediction arises from context-based inference. Formally, \( (q, t)\) should satisfy:
    \begin{equation}
     H(t \mid q) \geq \delta,
    \end{equation}
    where \( H(t \mid q) \) denotes the conditional entropy, and \( \delta\) is a threshold indicating a high level of uncertainty. \revise{The rationale is as follows: when $q$ already provides strong clues about the target $t$, $H(t \mid q)$ is low, and a hit may be inferred without memorization. In contrast, when $H(t \mid q)$ is high, $q$ conveys limited information about $t$, and a hit therefore constitutes stronger evidence of memorization. Nevertheless, this remains a probabilistic signal rather than proof of any individual hit: the model may still recover $t$ due to entropy-estimation error. We therefore rely on multiple rare and high-entropy hits instead of a single hit. The ablation study in Section~\ref{sec:ablation} further supports this design, showing that it effectively decreases the FPR.}
\end{packeditemize}

\noindent {\bf Hypothesis Testing.} We formalize the repository presence test \revise{as a standard binomial hypothesis-testing problem~\cite{farrington1990test}. This relies on the assumption that each signature snippet in $R$ constitutes an independent Bernoulli trial (hit or miss), thereby inducing a binomial distribution over the total number of hits observed by $M$. In practice, this assumption is only approximate: snippets from the same repository often share code structures, introducing cross-trial correlations that violate independence. To mitigate this issue, we apply both structural deduplication of code patterns and syntactic deduplication of query--target instances (Section~\ref{sec:repo_implement}).} 

\noindent The goal of hypothesis testing is to distinguish between the following two hypotheses:

\begin{packeditemize}
    \item \( H_0 \): a signature snippet's hit probability \(P\)  satisfies \(P \leq P_0\).
    \item \( H_1 \): a signature snippet's hit probability \(P\) satisfies \(P > P_0\).
\end{packeditemize}
where $P_0$ denotes the default hit probability when $R$ is not contained in $D$, i.e., 
\begin{equation} 
P_0 := P( t \in \text{Resp}(M; q) | R \notin D).
\end{equation}
\noindent \revise{\(P_0\) captures estimation errors in the signature-snippet conditions, along with stochastic hits from the model's random guessing.} In practice, \( P_0 \) can be estimated as the average hit-to-query ratio of \( M \), evaluated on repositories outside the training set \( D \) when feasible, or as a small fixed empirical value, such as 0.1. Again, \(\text{Resp}(M; q)\) denotes the set of responses generated by model $M$ when queried with $q$. 

Under \( H_0 \), the \textit{p}-value is the probability of observing at least \( k \) hits out of \( n \) trials and is computed as:
\begin{equation}
\label{eq:p-value}
\textit{p} := \sum_{i=k}^{n} \binom{n}{i} \cdot P_0^i \cdot (1 - P_0)^{n - i}.
\end{equation}

\noindent A $p$-value below the significance level allows for the rejection of the null hypothesis $H_0$, indicating that the hit frequency of model $M$ on repository $R$ significantly exceeds random expectations. In practice, we use the $p$-value directly as a repository-presence evidence score, where smaller values indicate stronger evidence of inclusion, and employ a weighted hit counting mechanism to account for distinctiveness and context-independence. Specifically, the weight for each hit is calculated as the ratio of the query's prediction entropy to the frequency of repositories containing the target word. This formulation is based on the rationale that successful predictions of rare targets with limited contextual information serve as strong indicators of memorization. Consequently, this approach removes the need for explicit feature thresholds and retains the default hit probability as the sole hyperparameter.

\noindent {\bf Challenges.} 
The goal of this stage is to identify as many signature snippets as possible within \( R \). However, several technical challenges must be addressed.

\noindent \textit{Challenge 1: 
Which repetitive patterns in \( R \) are most likely to be memorized by \( M \)?}

We extract frequently occurring patterns, either complete code snippets or their subsequences, from the repository \(R\). To capture the broader memorization behavior of transformer-based models, our pattern identification extends beyond contiguous sequences to include non-contiguous subsequences, since attention mechanisms can learn long-range token dependencies\revise{~\cite{vaswani2017attention}}. Consequently, even non-adjacent \revise{$(q,t)$} pairs may be memorized if they co-occur frequently in similar contexts. Details of the pattern types and extraction process are provided in \revise{Section~\ref{sec:repo_implement}}.

\noindent \textit{Challenge 2: How can we determine whether a target $t$ selected from a signature snippet is distinctive,  
i.e., unlikely to appear commonly across many repositories?}

A straightforward approach is to retain a target only if it occurs infrequently in a large reference corpus of repositories. Because code LLMs are commonly trained on public code such as GitHub repositories~\cite{github}, we use \textit{OctoPack}~\cite{muennighoff2023octopack} as a reference dataset\revise{. It is a reliable proxy: its common-word set has 98\% overlap with the set derived from The Stack v2.}

\revise{Furthermore, using lexical-level targets for distinctiveness determination is a scalability trade-off: while length-$l$ targets contain more unique structural information, verifying their global rarity would require matching every length-$l$ subsequence across all $N$ reference files (average length $L$), yielding $O(N \times (L + l))$ time and becoming infeasible at scale. In contrast, by simplifying the target $t$ to a single word ($l=1$), we can pre-compute a word-frequency dictionary over the reference dataset, allowing constant-time, $O(1)$, lookups.} This simplification not only improves computational efficiency but also enhances sensitivity: shorter targets are more likely to be recalled by the model, while longer sequences introduce greater output uncertainty and reduce the likelihood of exact matches.

\noindent \textit{Challenge 3: How can we quantify the information contained in a query $q$ for predicting a target $t$ while ensuring context independence?}

A straightforward approach is to use the loss from a reference model when predicting $t$ given $q$. However, this method faces two key issues:  
(1) The reference model may have been trained on $R$, leading to memorization bias and artificially low loss values; and  
(2) a target token $t$ often consists of sub-tokens that are either already present in the context or semantically predictable, causing the total loss to appear low even when $t$ itself is distinctive.

To mitigate the first issue, we employ a small-scale reference model, as smaller models exhibit weaker memorization and depend more on genuine inference~\cite{carlini2021extracting}. To address the second issue, we process the target $t$ at the sub-token level. Specifically, we first perform lexical analysis~\cite{hanks2013lexical} to identify sub-tokens in $t$ that have identical or lexically related counterparts in $q$ and exclude them from the loss computation. Next, we iteratively append each remaining sub-token to $q$ and compute the reference model’s loss for the unrevealed sub-tokens in $t$. Finally, we take the maximum loss across all sub-token evaluations since the conditional information entropy is best reflected by the hardest-to-predict sub-token.

\subsection{Stage II: Patch Presence Test}

Unlike typical patch-presence testing~\cite{zhang2018precise}, our approach is straightforward due to three key aspects of our application scenario. First, unlike conventional binary-level approaches~\cite{he2024bingo}, we have access to the source code. Second, while traditional methods target independent vulnerability patches, our bugs are localized within the target repository. Third, we leverage test suites provided by benchmarks, whereas conventional methods typically assume their absence.

\noindent \textbf{Patch Preprocessing.} We preprocess each patch to remove non-fixing elements such as comments and documentation, and extract the patch code blocks that implement the fix logic. Meanwhile, we normalize the patched code files to eliminate matching errors caused by differences in formatting and coding style.

\noindent \textbf{Patch Localization.} Naive full-text search is inefficient and prone to false positives due to internal code reuse~\cite{zhang2018precise}. To address this, \revise{we develop} a structured patch localization algorithm. First, we restrict the fixing scope to the specific function within the target class. Second, we employ an AST-based method to syntactically align the patch code with the snapshot code within this scope. Finally, when AST-based alignment fails, we utilize fuzzy syntax matching within the designated scope as a fallback mechanism.

\noindent \textbf{Similarity Matching.} For the aligned code block pairs, we compute similarity using a deep learning-based code embedding model. Compared to traditional AST-based similarity metrics, SOTA code embedding models can better capture semantic equivalence~\cite{kryvosheieva2025efficient}. If every patch block has a corresponding block in the cutoff snapshot with a similarity score above a predefined threshold, we conclude that the patch is already present in the cutoff snapshot; thus, the sample is contaminated.

\noindent \textbf{Testing Verification.} When similarity matching proves inconclusive, we employ the benchmark's test suite for final verification by generating a new diff file that replaces the patch with its aligned snapshot code. If this diff applies successfully to the buggy version and the resulting program passes all tests, the sample is deemed contaminated. This approach accounts for cases where extensive refactoring or the removal of non-essential logic obscures syntactic similarity. However, we use this process only as a fallback measure, as contextual changes in the snapshot (e.g., the introduction of new variables) frequently prevent the generated diff from applying even when the patch is genuinely present.

\section{System Implementation}

\subsection{Repository Presence Test} 
\label{sec:repo_implement}

\noindent {\bf Pattern Identification.} Specifically, we consider three forms of syntactic patterns: code block clones, text duplications, and word co-occurrences. 

{\textit{Code Block Clones.}} Code clones refer to identical or similar code blocks appearing across repository files due to reuse. To detect them, we employ NiCad 6.2~\cite{nicad6.2}, a text-based tool supporting multiple languages, including C, Java, and Python. NiCad groups the detected clones by type and outputs the results in an XML report, which we parse for analysis.

\textit{Text Duplications.} Text duplication refers to repeated auxiliary content in code files, such as copyright licenses, code comments, and import statements. 
While these text elements typically do not affect program execution logic, they often carry repository-specific contextual knowledge. To detect such text duplication patterns, we utilize the Simhash algorithm~\cite{sadowski2007simhash}. 
Specifically, we leverage Python's \texttt{SimHash}~\cite{python-simhash} library to compute text similarity, as described in Algorithm~\ref{alg:simhash_clustering} (Pseudocode for all algorithms is presented in Appendix~\ref{sec:appendix_a}). 
First, an empty dictionary is initialized to store the clustering results, where each key represents the hash value of a cluster, and each value is the set of texts within that cluster. Next, we iterate through the list of texts, compute the hash value of each text, and compare it against the hash values of existing clusters using the Hamming distance. 
If the distance is less than or equal to the predefined threshold, the text is added to the matching cluster; otherwise, a new cluster is created with the hash value as the key. Finally, the values of the dictionary are converted into a list and returned as the clustering result.

{\textit{Word Co-occurrences.}} Drawing on concepts from natural language processing~\cite{harris1957co}, we define word co-occurrence as the frequent appearance of multiple keywords across different repository files. Just as linguistic co-occurrence suggests semantic association, co-occurring keywords in code repositories indicate a form of contextual consistency. For instance, in Figure~\ref{fig:architecture}, \texttt{getOptions} and \texttt{enableTypeCheck} repeatedly appear in the same context, forming a co-occurrence pattern that indicates when the latter is invoked with the former. We formulate pattern detection as a frequent-pattern mining problem~\cite{han2007frequent} and employ the PrefixSpan algorithm~\cite{pei2004mining} for implementation. Algorithm~\ref{alg:pattern_extraction} outlines the detection process. First, each file is transformed into a sequence of distinctive words, denoted $\{w_1, \dots, w_m\}$; we then compute word and co-occurrence frequencies. We then construct a list of frequent sequences to capture file-level structures and feed it into the PrefixSpan algorithm~\cite{pei2001prefixspan} to extract raw patterns. Finally, these patterns are refined by merging those with identical frequencies and subset relationships to yield the final set of word co-occurrence patterns.

\noindent \textbf{Template Extraction.} As shown in Figure~\ref{fig:architecture}, \revise{we extract templates to deduplicate repeated patterns by abstracting their common content, thereby mitigating structural-level correlation effects between queries. Specifically,} each template is designed to be a structural pattern interleaved with \texttt{<text>} labels, which can be substituted with arbitrary content. 
We use Python's \texttt{difflib.SequenceMatcher}~\cite{python-difflib} to extract these templates from code block clone and text duplication patterns. Notably, word co-occurrence patterns can be regarded as implicit templates and do not require additional extraction. The SequenceMatcher algorithm determines similarity by identifying the longest common subsequence~\cite{hirschberg1977algorithms} between two sequences. 
The algorithm recursively divides the sequences, marks the matching segments, and processes the remaining unmatched parts to compute the final similarity score.

\begin{table*}[t]
\centering
\setlength{\tabcolsep}{3pt}        %
\renewcommand{\arraystretch}{0.9}  %
\caption{Statistics of six bug-fixing benchmarks and their overlap with The Stack v2.}
\label{tab:bugfix-stats}
\footnotesize
\begin{tabular*}{\textwidth}{@{\extracolsep{\fill}}l|ccc|ccc}
\toprule
\textbf{Benchmark} &
\textbf{SWE-bench Verified} &
\textbf{BugsInPy} &
\textbf{BugSwarm} &
\textbf{Multi-SWE-bench} &
\textbf{Defects4J} &
\textbf{GrowingBugs} \\
\midrule
Number of Repositories & 12 & 17 & 34 & 9 & 17 & 171 \\
Number of Bug-Fixing Samples & 500 & 500 & 170 & 128 & 835 & 1839 \\
Ratio of \revise{Repositories} Present in The Stack v2 & 100.0\% & 100.0\% & 85.2\% & 100.0\% & 100.0\% & 97.8\% \\
Ratio of \revise{Patches} Present in The Stack v2 & 78.9\% (3.2\%)  & 67.0\% (3.3\%) & 77.0\% (3.5\%) & 62.4\% (10.8\%) & 71.8\% (10.2\%) & 78.0\% (11.6\%) \\ 

\bottomrule
\end{tabular*}
\end{table*}

\noindent \textbf{Entropy Calculation.} The calculation of entropy to predict the target given a query is detailed in Algorithm~\ref{alg:loss_based_filtering}.
Specifically, both the target and the query are first tokenized into sub-token sequences. 
A query queue is maintained to construct updated contexts. 
For each sub-token in the target, if it either exactly matches any sub-token in the query or shares a lexical relation (e.g., stemming or synonymy), it is concatenated with the current query and enqueued.
Otherwise, the current query is dequeued as input, and the current sub-token is treated as the prediction target. This pair is then passed to a reference language model (i.e., {\em Qwen2.5-Coder-0.5B}~\cite{hui2024qwen2}) to compute the language modeling loss. 
After iterating over all sub-tokens in the target, the maximum loss across all sub-token predictions is used as the final entropy score.

\revise{\noindent \textbf{Syntactic Query Deduplication.} We further remove near-duplicate queries using a SequenceMatcher-style similarity check to mitigate syntactic correlations between queries. Specifically, each newly instantiated pair is compared with previously accepted query prefixes and discarded when it is near-duplicate.} 

\subsection{Patch Presence Test}

We use Python's \texttt{Black} library~\cite{black} and Java's \texttt{JavaParser} library~\cite{javaparser} for preprocessing patch code, including comment removal and formatting. The AST matching method we employ is GumTree~\cite{falleri2014fine}.

\noindent {\bf AST-Based Patch Localization.} As shown in Algorithm~\ref{alg:patch_block_locator}, the process of patch localization is as follows. First, we parse the patch diff file to extract the added code lines and construct an AST for them, while simultaneously parsing the target file in the repository to generate its complete AST. Second, we filter candidate functions from the repository AST as search scopes based on contextual features, such as function calls and variable names. Third, within each candidate function's AST subtree, we perform subtree matching using GumTree's tree edit distance algorithm to compute structural similarity and locate the exact position of the patch code block in the cutoff snapshot.

\revise{\noindent {\bf Embedding-Based Patch Matching.} For each patch logical line, we first perform exact matching. If no exact match is found, we use edit distance to retrieve candidate lines and then use embedding similarity for re-ranking. Finally, we count the lines whose similarity exceeds the threshold; only when all matched patch logical lines pass the threshold is the patch considered present in the cutoff snapshot. Specifically, we use \textit{jina-code-embeddings-0.5B}~\cite{kryvosheieva2025efficient}, a state-of-the-art code embedding model.}

\section{Experiment and Evaluation}

\subsection{Experimental Setup}
\label{sec:exp_setup}

\noindent \textbf{Environment Settings}. All experiments are conducted on a Linux (Red Hat Enterprise Linux 8.9) server, which is equipped with an AMD EPYC 7742 64-Core CPU, 500 GB of RAM, and 4 NVIDIA A100-80G GPUs.

\noindent \textbf{Evaluation Metrics}. We adopt the Area Under the Receiver Operating Characteristic Curve (AUC), True Positive Rate (TPR), and False Positive Rate (FPR) as the evaluation metrics, which are also used in prior work~\cite{carlini2021extracting, duan2024membership}. 

\newcolumntype{Y}{>{\centering\arraybackslash}X}

\begin{table*}[t]
\centering
\footnotesize
\setlength{\tabcolsep}{2pt}
\renewcommand{\arraystretch}{1.03}

\newcommand{\benchsize}{\footnotesize}       %
\newcommand{\modelsize}{\footnotesize}%

\begin{threeparttable}
\caption{Evaluation of methods \revise{for detecting contaminated samples in LLM bug-fixing benchmarks.}}
\label{tab:benchmarks_models}

\begin{tabularx}{\textwidth}{@{\extracolsep{\fill}} ll|*{3}{Y}|*{3}{Y}|*{3}{Y}|*{3}{Y}|*{3}{Y}|*{3}{Y}}
\toprule
\multicolumn{2}{l|}{\textbf{\benchsize Benchmark}} &
\multicolumn{3}{c|}{\benchsize \textbf{SWE-bench Verified}} &
\multicolumn{3}{c|}{\benchsize \textbf{BugsInPy}} &
\multicolumn{3}{c|}{\benchsize \textbf{BugSwarm}} &
\multicolumn{3}{c|}{\benchsize \textbf{Multi-SWE-bench}} &
\multicolumn{3}{c|}{\benchsize \textbf{Defects4J}} &
\multicolumn{3}{c}{\benchsize \textbf{GrowingBugs}} \\
\cmidrule(lr){1-2}\cmidrule(lr){3-5}\cmidrule(lr){6-8}\cmidrule(lr){9-11}\cmidrule(lr){12-14}\cmidrule(lr){15-17}\cmidrule(lr){18-20}
\multicolumn{2}{l|}{\textbf{\modelsize Model}} &
\modelsize SC & \modelsize SC2 & \modelsize CG2 &
\modelsize SC & \modelsize SC2 & \modelsize CG2 &
\modelsize SC & \modelsize SC2 & \modelsize CG2 &
\modelsize SC & \modelsize SC2 & \modelsize CG2 &
\modelsize SC & \modelsize SC2 & \modelsize CG2 &
\modelsize SC & \modelsize SC2 & \modelsize CG2 \\
\midrule

\multirow{3}{*}{\textbf{LOSS}} & AUC & 0.486 & 0.465 & 0.480 & 0.465 & 0.466 & 0.479 & 0.488 & 0.485 & 0.492 & 0.509 & 0.583 & 0.482 & 0.610 & 0.591 & 0.461 & 0.490 & 0.490 & 0.492 \\
& TPR & 0.916 & 0.889 & 0.164 & 0.164 & 0.529 & 0.920 & 0.907 & 0.919 & 0.709 & 0.333 & 0.333 & 0.333 & 0.636 & 0.591 & 0.227 & 0.175 & 0.143 & 0.082 \\
& FPR & 0.872 & 0.859 & 0.128 & 0.131 & 0.514 & 0.891 & 0.756 & 0.778 & 0.578 & 0.000 & 0.000 & 0.111 & 0.385 & 0.346 & 0.115 & 0.128 & 0.102 & 0.051 \\
\midrule
\multirow{3}{*}{\textbf{Zlib}} & AUC & 0.506 & 0.494 & 0.498 & 0.460 & 0.461 & 0.478 & 0.544 & 0.546 & 0.534 & 0.463 & 0.509 & 0.417 & 0.584 & 0.575 & 0.478 & 0.507 & 0.507 & 0.504 \\
& TPR & 0.942 & 0.865 & 0.871 & 0.933 & 0.908 & 0.954 & 0.837 & 0.837 & 0.430 & 0.167 & 0.667 & 0.083 & 0.614 & 0.318 & 0.205 & 0.181 & 0.216 & 0.116 \\
& FPR & 0.872 & 0.795 & 0.808 & 0.902 & 0.880 & 0.929 & 0.644 & 0.667 & 0.289 & 0.000 & 0.444 & 0.000 & 0.385 & 0.115 & 0.115 & 0.125 & 0.158 & 0.066 \\
\midrule
\multirow{3}{*}{\textbf{Min-K\%}} & AUC & 0.524 & 0.504 & 0.491 & 0.462 & 0.468 & 0.476 & 0.543 & 0.566 & 0.466 & 0.472 & 0.602 & 0.482 & 0.601 & 0.609 & 0.434 & 0.494 & 0.499 & 0.487 \\
& TPR & 0.383 & 0.443 & 0.143 & 0.029 & 0.197 & 0.122 & 0.628 & 0.593 & 0.174 & 0.167 & 0.667 & 0.250 & 0.818 & 0.477 & 0.205 & 0.227 & 0.204 & 0.111 \\
& FPR & 0.295 & 0.346 & 0.090 & 0.011 & 0.180 & 0.082 & 0.467 & 0.333 & 0.022 & 0.000 & 0.333 & 0.000 & 0.577 & 0.231 & 0.154 & 0.182 & 0.128 & 0.084 \\
\midrule
\multirow{3}{*}{\textbf{Ref}} & AUC & 0.535 & 0.449 & 0.524 & 0.539 & 0.491 & 0.473 & 0.554 & 0.576 & 0.654 & 0.602 & 0.583 & 0.407 & 0.517 & 0.546 & 0.519 & 0.511 & 0.519 & 0.473 \\
& TPR & 0.894 & 0.190 & 0.435 & 0.319 & 0.580 & 0.618 & 0.512 & 0.302 & 0.570 & 0.917 & 0.583 & 0.583 & 0.477 & 0.477 & 0.432 & 0.850 & 0.527 & 0.110 \\
& FPR & 0.795 & 0.167 & 0.333 & 0.235 & 0.525 & 0.568 & 0.267 & 0.000 & 0.156 & 0.667 & 0.222 & 0.444 & 0.346 & 0.269 & 0.231 & 0.812 & 0.466 & 0.081 \\
\midrule

\multirow{3}{*}{\textbf{Ours}} & AUC & \textbf{0.945} & \textbf{0.945} & \textbf{0.945} & \textbf{0.909} & \textbf{0.919} & \textbf{0.905} & \textbf{0.865} & \textbf{0.934} & \textbf{0.875} & \textbf{0.919} & \textbf{0.919} & \textbf{0.919} & \textbf{0.863} & \textbf{0.850} & \textbf{0.863} & \textbf{0.863} & \textbf{0.838} & \textbf{0.859} \\
& TPR & \textbf{0.979} & \textbf{0.979} & \textbf{0.979} & \textbf{0.872} & \textbf{0.893} & \textbf{0.864} & \textbf{0.816} & \textbf{0.954} & \textbf{0.816} & \textbf{0.938} & \textbf{0.938} & \textbf{0.938} & \textbf{0.777} & \textbf{0.746} & \textbf{0.777} & \textbf{0.795} & \textbf{0.753} & \textbf{0.792} \\
& FPR & \textbf{0.090} & \textbf{0.090} & \textbf{0.090} & \textbf{0.055} & \textbf{0.055} & \textbf{0.055} & \textbf{0.087} & \textbf{0.087} & \textbf{0.065} & \textbf{0.100} & \textbf{0.100} & \textbf{0.100} & \textbf{0.051} & \textbf{0.046} & \textbf{0.051} & \textbf{0.069} & \textbf{0.078} & \textbf{0.075} \\

\bottomrule
\end{tabularx}

\begin{tablenotes}[para,flushleft]
\footnotesize SC : StarCoder~\cite{li2023starcoder}; SC2 : StarCoder2~\cite{lozhkov2024starcoder}; CG2 : CodeGen2~\cite{nijkamp2023codegen2}.
\end{tablenotes}
\end{threeparttable}
\end{table*}

\noindent \textbf{Baselines}. To the best of our knowledge, \TN{} is the first method designed to detect contaminated samples in bug-fixing benchmarks. For comparison, we evaluate four widely used logit-based membership inference baselines that are currently applied to data contamination detection~\cite{kong2025demystifying}: {Loss}~\cite{yeom2018privacy}, {Zlib}~\cite{carlini2021extracting}, {Reference-based method (Ref)}~\cite{carlini2021extracting}, and {Min-$k$\%}~\cite{shi2023detecting}. 

\subsection{Dataset and Benchmark}
\label{sec:datasets}

\noindent\textbf{Dataset and Model.} To validate \TN{}'s effectiveness, \revise{we use models and datasets for which sample membership can be verified. Specifically, we use The Stack v2~\cite{lozhkov2024starcoder} as the reference training dataset}. It is the largest publicly available code-repository training set to date, containing 7,505,725 Python repositories and 7,997,020 Java repositories, with a cutoff date of September 2023. \revise{We use three evaluation models trained on The Stack v2:} StarCoder-15B~\cite{li2023starcoder}, StarCoder2-15B~\cite{lozhkov2024starcoder}, and CodeGen2-16B~\cite{nijkamp2023codegen2}.

\noindent\textbf{Benchmarks.} We select six widely used bug-fixing benchmarks in Python and Java for evaluation: SWE-bench Verified~\cite{chowdhury2024swebenchverified}, BugsInPy~\cite{widyasari2020bugsinpy}, and BugSwarm~\cite{tomassi2019bugswarm} for Python; Multi-SWE-bench~\cite{zan2025multi}, Defects4J~\cite{just2014defects4j}, and GrowingBugs~\cite{zhu2023reproducibility} for Java. 

\noindent\textbf{Sample Analysis.} We manually inspect the presence of samples in The Stack v2 and analyze the contamination overlap for each benchmark. As shown in Table~\ref{tab:bugfix-stats}, approximately 70\% of benchmark patches, and over 85\% of their source repositories, are present in The Stack v2, indicating a serious data contamination issue. Furthermore, \revise{we analyze the prevalence of Category~\Circled{3} in Figure~\ref{fig:timeline}}, specifically patches that are absent from the cutoff snapshot but persist in The Stack v2, as detailed in parentheses in Table~\ref{tab:bugfix-stats}. The results indicate that such samples constitute approximately 3\% of the Python benchmark, whereas the proportion is notably higher in Java, reaching around 10\%. We attribute this to the collection of outdated repository versions through code reuse (clone or fork), with a higher prevalence within the Java ecosystem compared to Python~\cite{lopes2017dejavu}.

\subsection{Performance Evaluation}

\revise{This section evaluates both the complete two-stage performance of \TN{} and the standalone performance of each stage. For Stage~I, we isolate repository-inclusion detection by labeling benchmark repositories according to their The Stack v2 membership and adding post-cutoff repositories as negatives. For Stage~II, we remove the influence of Stage~I by assuming that benchmark repositories are already present in The Stack v2, allowing us to evaluate patch-presence detection directly.}

\begin{table*}[t]
\centering
\footnotesize
\setlength{\tabcolsep}{2pt}
\renewcommand{\arraystretch}{1.03}

\newcommand{\benchsize}{\footnotesize}       %
\newcommand{\modelsize}{\footnotesize}%

\begin{threeparttable}
\caption{Evaluation of methods \revise{for detecting repository presence} in the LLMs' training set across 6 benchmarks.}
\label{tab:stage_1_eval}

\begin{tabularx}{\textwidth}{@{\extracolsep{\fill}} ll|*{3}{Y}|*{3}{Y}|*{3}{Y}|*{3}{Y}|*{3}{Y}|*{3}{Y}}
\toprule
\multicolumn{2}{l|}{\textbf{\benchsize Benchmark}} &
\multicolumn{3}{c|}{\benchsize \textbf{SWE-bench Verified}} &
\multicolumn{3}{c|}{\benchsize \textbf{BugsInPy}} &
\multicolumn{3}{c|}{\benchsize \textbf{BugSwarm}} &
\multicolumn{3}{c|}{\benchsize \textbf{Multi-SWE-bench}} &
\multicolumn{3}{c|}{\benchsize \textbf{Defects4J}} &
\multicolumn{3}{c}{\benchsize \textbf{GrowingBugs}} \\
\cmidrule(lr){1-2}\cmidrule(lr){3-5}\cmidrule(lr){6-8}\cmidrule(lr){9-11}\cmidrule(lr){12-14}\cmidrule(lr){15-17}\cmidrule(lr){18-20}
\multicolumn{2}{l|}{\textbf{\modelsize Model}} &
\modelsize SC & \modelsize SC2 & \modelsize CG2 &
\modelsize SC & \modelsize SC2 & \modelsize CG2 &
\modelsize SC & \modelsize SC2 & \modelsize CG2 &
\modelsize SC & \modelsize SC2 & \modelsize CG2 &
\modelsize SC & \modelsize SC2 & \modelsize CG2 &
\modelsize SC & \modelsize SC2 & \modelsize CG2 \\
\midrule

\multirow{3}{*}{\textbf{LOSS}} & AUC & 0.789 & 0.765 & 0.765 & 0.786 & 0.715 & 0.758 & 0.747 & 0.697 & 0.734 & 0.927 & 0.822 & 0.915 & 0.927 & 0.851 & 0.908 & 0.911 & 0.850 & 0.881 \\
& TPR & 0.818 & 0.727 & 0.818 & 0.882 & 0.706 & 0.824 & 0.536 & 0.676 & 0.643 & 1.000 & 0.857 & 0.857 & 0.824 & 0.824 & 0.824 & 0.802 & 0.737 & 0.778 \\
& FPR & 0.190 & 0.120 & 0.290 & 0.340 & 0.308 & 0.330 & 0.110 & 0.316 & 0.210 & 0.221 & 0.233 & 0.081 & 0.093 & 0.233 & 0.081 & 0.093 & 0.151 & 0.105 \\
\midrule
\multirow{3}{*}{\textbf{Zlib}} & AUC & 0.485 & 0.507 & 0.474 & 0.497 & 0.507 & 0.494 & 0.560 & 0.527 & 0.555 & 0.852 & 0.806 & 0.841 & 0.899 & 0.865 & 0.882 & 0.834 & 0.800 & 0.819 \\
& TPR & 0.636 & 0.545 & 0.636 & 0.882 & 0.765 & 0.824 & 0.750 & 0.750 & 0.786 & 1.000 & 1.000 & 1.000 & 0.941 & 0.941 & 0.941 & 0.766 & 0.719 & 0.659 \\
& FPR & 0.420 & 0.316 & 0.430 & 0.650 & 0.581 & 0.620 & 0.510 & 0.607 & 0.560 & 0.221 & 0.337 & 0.267 & 0.221 & 0.337 & 0.267 & 0.186 & 0.186 & 0.105 \\
\midrule
\multirow{3}{*}{\textbf{Min-K\%}} & AUC & 0.847 & 0.807 & 0.837 & 0.841 & 0.737 & 0.804 & 0.779 & 0.724 & 0.754 & 0.935 & 0.846 & 0.930 & 0.936 & 0.875 & 0.927 & 0.920 & 0.866 & 0.899 \\
& TPR & 0.818 & 0.727 & 0.818 & 0.882 & 0.706 & 0.706 & 0.750 & 0.741 & 0.679 & 1.000 & 1.000 & 0.857 & 1.000 & 0.824 & 0.824 & 0.743 & 0.719 & 0.784 \\
& FPR & 0.110 & 0.060 & 0.240 & 0.280 & 0.299 & 0.220 & 0.300 & 0.308 & 0.220 & 0.209 & 0.372 & 0.070 & 0.209 & 0.186 & 0.070 & 0.047 & 0.093 & 0.081 \\
\midrule
\multirow{3}{*}{\textbf{Ref}} & AUC & 0.988 & 0.961 & 0.605 & 0.962 & 0.951 & 0.592 & 0.856 & 0.886 & 0.597 & 0.939 & 0.762 & 0.915 & 0.965 & 0.878 & 0.924 & 0.888 & 0.736 & 0.876 \\
& TPR & 1.000 & 1.000 & 0.636 & 0.941 & 0.941 & 0.765 & 0.964 & 0.898 & 0.393 & 0.857 & 0.714 & 1.000 & 1.000 & 1.000 & 1.000 & 0.898 & 0.671 & 0.719 \\
& FPR & 0.080 & 0.154 & 0.280 & 0.050 & 0.120 & 0.540 & 0.180 & 0.231 & 0.170 & 0.058 & 0.209 & 0.174 & 0.140 & 0.244 & 0.244 & 0.186 & 0.256 & 0.181 \\
\midrule

\multirow{3}{*}{\textbf{Ours}} & AUC & \textbf{1.000} & \textbf{0.999} & \textbf{1.000} & \textbf{0.978} & \textbf{0.972} & \textbf{0.969} & \textbf{0.883} & \textbf{0.895} & \textbf{0.892} & \textbf{0.985} & \textbf{0.925} & \textbf{0.991} & \textbf{0.982} & \textbf{0.966} & \textbf{0.990} & \textbf{0.912} & \textbf{0.863} & \textbf{0.908} \\
& TPR & \textbf{1.000} & \textbf{1.000} & \textbf{1.000} & \textbf{0.941} & \textbf{1.000} & \textbf{0.882} & \textbf{0.786} & \textbf{0.964} & \textbf{0.821} & \textbf{1.000} & \textbf{1.000} & \textbf{1.000} & \textbf{1.000} & \textbf{0.882} & \textbf{1.000} & \textbf{0.790} & \textbf{0.802} & \textbf{0.832} \\
& FPR & \textbf{0.000} & \textbf{0.010} & \textbf{0.000} & \textbf{0.040} & \textbf{0.170} & \textbf{0.050} & \textbf{0.080} & \textbf{0.290} & \textbf{0.080} & \textbf{0.060} & \textbf{0.298} & \textbf{0.036} & \textbf{0.095} & \textbf{0.071} & \textbf{0.060} & \textbf{0.119} & \textbf{0.238} & \textbf{0.167} \\

\bottomrule
\end{tabularx}
\begin{tablenotes}[para,flushleft]
\footnotesize SC : StarCoder~\cite{li2023starcoder}; SC2 : StarCoder2~\cite{lozhkov2024starcoder}; CG2 : CodeGen2~\cite{nijkamp2023codegen2}.
\end{tablenotes}
\end{threeparttable}
\vspace{-0.45em}
\end{table*}

\begin{table*}[t]
\centering
\footnotesize
\setlength{\tabcolsep}{2pt}
\renewcommand{\arraystretch}{1.03}

\newcommand{\benchsize}{\footnotesize}       %
\newcommand{\modelsize}{\footnotesize}%

\begin{threeparttable}
\caption{Evaluation of methods \revise{for detecting patch presence when the repository} is included in the LLMs' training set.}
\label{tab:stage_2_eval}

\begin{tabularx}{\textwidth}{@{\extracolsep{\fill}} ll|*{3}{Y}|*{3}{Y}|*{3}{Y}|*{3}{Y}|*{3}{Y}|*{3}{Y}}
\toprule
\multicolumn{2}{l|}{\textbf{\benchsize Benchmark}} &
\multicolumn{3}{c|}{\benchsize \textbf{SWE-bench Verified}} &
\multicolumn{3}{c|}{\benchsize \textbf{BugsInPy}} &
\multicolumn{3}{c|}{\benchsize \textbf{BugSwarm}} &
\multicolumn{3}{c|}{\benchsize \textbf{Multi-SWE-bench}} &
\multicolumn{3}{c|}{\benchsize \textbf{Defects4J}} &
\multicolumn{3}{c}{\benchsize \textbf{GrowingBugs}} \\
\cmidrule(lr){1-2}\cmidrule(lr){3-5}\cmidrule(lr){6-8}\cmidrule(lr){9-11}\cmidrule(lr){12-14}\cmidrule(lr){15-17}\cmidrule(lr){18-20}
\multicolumn{2}{l|}{\textbf{\modelsize Model}} &
\modelsize SC & \modelsize SC2 & \modelsize CG2 &
\modelsize SC & \modelsize SC2 & \modelsize CG2 &
\modelsize SC & \modelsize SC2 & \modelsize CG2 &
\modelsize SC & \modelsize SC2 & \modelsize CG2 &
\modelsize SC & \modelsize SC2 & \modelsize CG2 &
\modelsize SC & \modelsize SC2 & \modelsize CG2 \\
\midrule

\multirow{3}{*}{\textbf{LOSS}} & AUC & 0.494 & 0.474 & 0.468 & 0.465 & 0.466 & 0.479 & 0.536 & 0.532 & 0.487 & 0.509 & 0.583 & 0.482 & 0.486 & 0.482 & 0.479 & 0.489 & 0.490 & 0.491 \\
& TPR & 0.858 & 0.889 & 0.164 & 0.164 & 0.529 & 0.920 & 0.837 & 0.919 & 0.209 & 0.333 & 0.333 & 0.333 & 0.649 & 0.162 & 0.096 & 0.175 & 0.143 & 0.082 \\
& FPR & 0.809 & 0.838 & 0.132 & 0.131 & 0.514 & 0.891 & 0.615 & 0.731 & 0.077 & 0.000 & 0.000 & 0.111 & 0.613 & 0.124 & 0.060 & 0.129 & 0.101 & 0.049 \\
\midrule
\multirow{3}{*}{\textbf{Zlib}} & AUC & 0.520 & 0.510 & 0.503 & 0.460 & 0.461 & 0.478 & 0.669 & 0.670 & 0.631 & 0.463 & 0.509 & 0.417 & 0.497 & 0.494 & 0.487 & 0.507 & 0.506 & 0.503 \\
& TPR & 0.829 & 0.865 & 0.905 & 0.933 & 0.908 & 0.954 & 0.500 & 0.512 & 0.430 & 0.167 & 0.667 & 0.083 & 0.280 & 0.265 & 0.171 & 0.181 & 0.216 & 0.116 \\
& FPR & 0.735 & 0.765 & 0.824 & 0.902 & 0.880 & 0.929 & 0.115 & 0.154 & 0.115 & 0.000 & 0.444 & 0.000 & 0.230 & 0.217 & 0.115 & 0.126 & 0.160 & 0.064 \\
\midrule
\multirow{3}{*}{\textbf{Min-K\%}} & AUC & 0.536 & 0.514 & 0.472 & 0.462 & 0.468 & 0.476 & 0.580 & 0.589 & 0.486 & 0.472 & 0.602 & 0.482 & 0.496 & 0.496 & 0.468 & 0.495 & 0.500 & 0.485 \\
& TPR & 0.383 & 0.443 & 0.143 & 0.029 & 0.198 & 0.122 & 0.628 & 0.593 & 0.233 & 0.167 & 0.667 & 0.250 & 0.137 & 0.220 & 0.068 & 0.134 & 0.204 & 0.111 \\
& FPR & 0.279 & 0.338 & 0.088 & 0.011 & 0.180 & 0.082 & 0.385 & 0.308 & 0.039 & 0.000 & 0.333 & 0.000 & 0.083 & 0.152 & 0.028 & 0.089 & 0.129 & 0.083 \\
\midrule
\multirow{3}{*}{\textbf{Ref}} & AUC & 0.539 & 0.448 & 0.513 & 0.539 & 0.491 & 0.473 & 0.545 & 0.567 & 0.678 & 0.602 & 0.583 & 0.407 & 0.524 & 0.530 & 0.489 & 0.514 & 0.523 & 0.475 \\
& TPR & 0.895 & 0.325 & 0.515 & 0.319 & 0.580 & 0.618 & 0.512 & 0.302 & 0.570 & 0.917 & 0.583 & 0.583 & 0.543 & 0.338 & 0.132 & 0.569 & 0.527 & 0.110 \\
& FPR & 0.779 & 0.294 & 0.427 & 0.235 & 0.525 & 0.568 & 0.308 & 0.000 & 0.077 & 0.667 & 0.222 & 0.444 & 0.479 & 0.244 & 0.092 & 0.525 & 0.457 & 0.080 \\
\midrule

\multirow{3}{*}{\textbf{Ours}} & AUC & \textbf{0.938} & \textbf{0.938} & \textbf{0.938} & \textbf{0.920} & \textbf{0.920} & \textbf{0.920} & \textbf{0.919} & \textbf{0.919} & \textbf{0.919} & \textbf{0.912} & \textbf{0.912} & \textbf{0.912} & \textbf{0.863} & \textbf{0.863} & \textbf{0.863} & \textbf{0.879} & \textbf{0.879} & \textbf{0.879} \\
& TPR & \textbf{0.979} & \textbf{0.979} & \textbf{0.979} & \textbf{0.893} & \textbf{0.893} & \textbf{0.893} & \textbf{0.954} & \textbf{0.954} & \textbf{0.954} & \textbf{0.938} & \textbf{0.938} & \textbf{0.938} & \textbf{0.777} & \textbf{0.777} & \textbf{0.777} & \textbf{0.850} & \textbf{0.850} & \textbf{0.850} \\
& FPR & \textbf{0.103} & \textbf{0.103} & \textbf{0.103} & \textbf{0.055} & \textbf{0.055} & \textbf{0.055} & \textbf{0.115} & \textbf{0.115} & \textbf{0.115} & \textbf{0.100} & \textbf{0.100} & \textbf{0.100} & \textbf{0.051} & \textbf{0.051} & \textbf{0.051} & \textbf{0.092} & \textbf{0.092} & \textbf{0.092} \\

\bottomrule
\end{tabularx}
\begin{tablenotes}[para,flushleft]
\footnotesize SC : StarCoder~\cite{li2023starcoder}; SC2 : StarCoder2~\cite{lozhkov2024starcoder}; CG2 : CodeGen2~\cite{nijkamp2023codegen2}.
\end{tablenotes}
\end{threeparttable}
\end{table*}

Table~\ref{tab:benchmarks_models} compares \TN{} with baselines in detecting contaminated samples across bug-fixing benchmarks. Using only textual \revise{outputs}, our method significantly outperforms \revise{logit-based} baselines across all six benchmarks and three models. Specifically, it achieves an AUC exceeding 0.9 on SWE-bench Verified, BugsInPy, and Multi-SWE-bench. In contrast, baselines typically yield AUC values around 0.5, indicating performance close to random guessing. \revise{Similar findings for these methods were reported} in~\cite{kong2025demystifying}. Notably, our method maintains an FPR at or below 0.1 in the complete evaluation by leveraging LLM training data collection practices. However, performance is relatively lower on Java benchmarks (Defects4J and GrowingBugs) due to prevalent code reuse, as discussed in Section~\ref{sec:datasets}, which increases false negatives and reduces TPR.

\noindent\textbf{Repository Presence Test Evaluation}. Table~\ref{tab:stage_1_eval} demonstrates that \TN{} consistently outperforms the baselines in the repository presence test. Notably, on SWE-bench Verified and Multi-SWE-bench, our method achieves near-perfect performance, with AUC and TPR values approaching 1.0 across all three models, indicating the successful detection of all training-included repositories. Note that most of the repositories in these benchmarks are included in The Stack v2, resulting in too few negative samples. To address the label imbalance issue, we expanded the negative set following Multi-SWE-bench standards~\cite{zan2025multi}. Specifically, we selected repositories created after September 2023 that have at least 500 stars and have built-in automated testing configurations. Consequently, we added 100 such repositories as negative samples for both the Python and Java benchmarks.

\revise{Stage I false positives are mainly caused by code reuse in post-cutoff negative repositories (90.3\%). Although these repositories are labeled as absent in our evaluation, many contain fragments reused from repositories collected for training sets before the cutoff; thus, these hit signals often reflect memorized reused code rather than accidental token prediction. For example, a misclassified repository \texttt{BizSpringSource/bizspring-vue3-opensource} reuses Spring-based business application scaffold code, triggering hits on rare identifiers such as \texttt{AuthorizationScope}.}

\Needspace{5\baselineskip}
\noindent\textbf{Patch Presence Test Evaluation}. Table~\ref{tab:stage_2_eval} presents the patch presence test results. While the baselines lack discriminative power, \TN{} maintains a high AUC of approximately 0.9. Notably, because \TN{} relies solely on the training cutoff date without querying models, it yields identical results for all models trained on The Stack v2. The failure of the baselines stems from their inability to determine whether fine-grained patch elements are memorized. For instance, Figure~\ref{lst:astropy__astropy-12907} depicts a single-line patch changing a literal value from \texttt{1} to \texttt{right}; baselines assign nearly identical scores before and after the change because the logit-based scores are insensitive to minor edits. Conversely, \TN{} successfully detects the patch by recognizing its unmodified presence in the cutoff snapshot. 

\revise{Stage~II false negatives mainly fall into two categories. The dominant category consists of patches that appear in earlier training snapshots but are no longer observable in the cutoff snapshot (87.0\%). This happens when older forks, releases, or historical revisions preserve patched code, but the inspected cutoff state has removed, renamed, or reorganized the relevant block; Stage~II then lacks an exact block to match. The second category consists of heavily refactored patches (13.0\%), where the fix is semantically preserved but moved across files, functions, or validation layers. For example, as shown in Figure~\ref{lst:JacksonDatabind_78}, this case moves the illegal-deserialization check from \texttt{BeanDeserializerFactory} into a separate \texttt{SubTypeValidator}, leaving only a wrapper call in the original file, so similarity matching captures the wrapper but not the original validation logic. We further study the robustness of detecting these refactored patch-present cases: similarity matching handles 77.0\% of moderate rewrites, test-suite verification recovers another 11.0\%, and the remaining 12.0\% remain challenging.}

\begin{figure}[t]
\begin{lstlisting}[style=arxivdiff]
---a/fasterxml/jackson/databind/deser/BeanDeserializerFactory.java
+++b/fasterxml/jackson/databind/deser/BeanDeserializerFactory.java
@@ -151,3 +139,3 @@ public JsonDeserializer<Object> createBeanDeserializer(...)
-        checkIllegalTypes(ctxt, type, beanDesc);
+        _validateSubType(ctxt, type, beanDesc);
@@ -873,17 +1025,9 @@ public BeanDeserializerFactory
-    protected void checkIllegalTypes(Context ctxt, JavaType type,
+    protected void _validateSubType(Context ctxt, JavaType type,
             BeanDescription beanDesc) throws JsonMappingException{
-        String full = type.getRawClass().getName();
-        if (_cfgIllegalClassNames.contains(full)) {
-            ctxt.reportBadTypeDefinition(beanDesc,
-                    "Illegal type (%s) to deserialize", full); }
+        SubTypeValidator.instance().validateSubType(ctxt, type, beanDesc);
\end{lstlisting}
\caption{\revise{Patch for the sample with instance ID \texttt{JacksonDatabind\_78} on Defects4J.}}
\label{lst:JacksonDatabind_78}
\end{figure}

\begin{figure}[t]
\begin{lstlisting}[style=arxivdiff]
--- a/astropy/modeling/separable.py
+++ b/astropy/modeling/separable.py
@@ -242,7 +242,7 @@ def _cstack(left, right):
    cright = _coord_matrix(right, 'right', noutp)
else:
    cright = np.zeros((noutp, right.shape[1]))
-   cright[-right.shape[0]:, -right.shape[1]:] = 1
+   cright[-right.shape[0]:, -right.shape[1]:] = right
 return np.hstack([cleft, cright])
\end{lstlisting}
\caption{\revise{Patch for the sample with instance ID \texttt{astropy\_\_astropy-12907} on SWE-bench Verified.}}
\label{lst:astropy__astropy-12907}
\end{figure}

\subsection{Ablation Study}
\label{sec:ablation}

We conduct the ablation study on the SWE-bench Verified dataset using StarCoder-15B, as shown in Table~\ref{tab:ablation}. 

\noindent\textbf{Repository Presence Test}. We evaluate three design criteria for signature snippet selection. First, disabling pattern identification to assess \textit{repetitiveness} results in a significant TPR drop to 0.428, indicating that models struggle to reproduce isolated distinctive words despite their presence in the training data. Second, evaluating \textit{distinctiveness} via unconstrained target selection leads to a substantial FPR increase, as common tokens are easily predicted regardless of training exposure. Third, examining \textit{context-independence} by removing entropy constraints causes a sharp FPR rise to 0.323, suggesting that contextually informative queries can lead to false positives.

\begin{table}[t]
\centering
\footnotesize
\caption{Ablation evaluation results for each stage.}
\label{tab:ablation}
\begin{tabular}{lccc}
\hline
\textbf{Stage I: Repository Presence Test} & \textbf{AUC} & \textbf{TPR} & \textbf{FPR}\\
\hline
\TN{}                                         & 1.000 & 1.000 & 0.000\\
\TN{} w/o Repetitiveness                      & 0.663 & 0.428 & 0.064\\
\TN{} w/o Distinctiveness                     & 0.647 & 0.871 & 0.486\\
\TN{} w/o Context-\revise{Independence}       & 0.845 & 0.952 & 0.323\\
\hline
\textbf{Stage II: Patch Presence Test} & \textbf{AUC} & \textbf{TPR} & \textbf{FPR}\\
\hline
\TN{}                                         & 0.938 & 0.979 & 0.103\\
\TN{} w/o Patch Preprocessing                & 0.786 & 0.762 & 0.135 \\
\TN{} w/o Patch Localization                  & 0.874 & 0.932 & 0.287 \\
\TN{} w/o Similarity Matching                 & 0.688 & 0.652 & 0.050 \\
\TN{} w/o Testing Verification                & 0.868 & 0.891 & 0.092 \\
\hline
\end{tabular}
\end{table}

\noindent\textbf{Patch Presence Test}. We evaluate the necessity of four processing steps. First, omitting formatting and comment removal lowers the AUC to 0.786, as noise from non-functional changes impairs TPR. Second, bypassing patch localization in favor of direct line-by-line comparison increases the FPR to 0.287, due to patch lines coincidentally appearing elsewhere in the file. Third, substituting similarity calculation with exact string matching causes a substantial AUC drop to 0.688, as strict matching fails to identify equivalent refactorings. Fourth, classifying below-threshold cases without test-suite verification decreases TPR because highly refactored equivalent patches are misclassified as clean.

\subsection{Hyperparameter Analysis}
\label{sec:threshold_analysis}

\revise{During evaluation, we split the hyperparameter-tuning subsets from BugSwarm (Python) and GrowingBugs (Java) into validation and test sets with a ratio of 1:4. The validation split is used for parameter selection, while the held-out split is used only for evaluation.}

\begin{figure}
    \centering
    \subfigure{\includegraphics[width=0.49\linewidth]{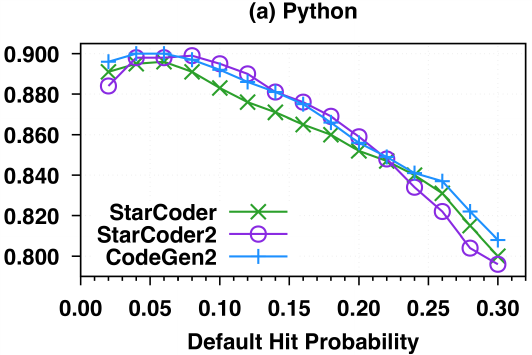}}
    \subfigure{\includegraphics[width=0.49\linewidth]{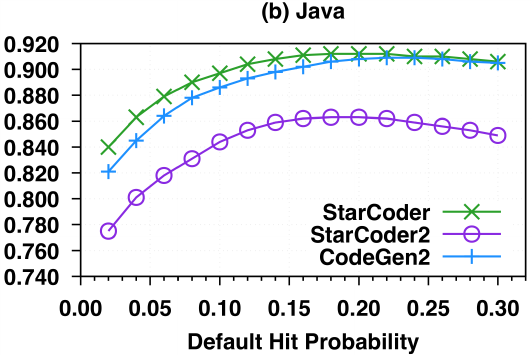}}
    \caption{AUC under different default hit probabilities.}
    \label{fig:default_p0}
    \vspace{-1.1em}
\end{figure}

\noindent\textbf{Default Hit Probability}. In the repository presence test, the parameter $P_0$ denotes the empirical probability of correctly predicting target suffixes from unseen signature snippets, which is used to compute the $p$-value in Eq.~\ref{eq:p-value}. Setting $P_0$ too low underestimates the model's capability on unseen snippets and increases false positives, whereas setting it too high compromises recall. Figure~\ref{fig:default_p0} depicts the impact of $P_0$ on the AUC for identifying included repositories in BugSwarm (Python) and GrowingBugs (Java). The results reveal consistent trends across models, where the optimal $P_0$ is approximately 0.08 for Python and 0.2 for Java. This disparity arises because Java exhibits stronger structural regularity and redundancy than Python, which allows for more robust prediction on unseen snippets.

\begin{figure}[t]
    \centering
    \subfigure{\includegraphics[width=0.49\linewidth]{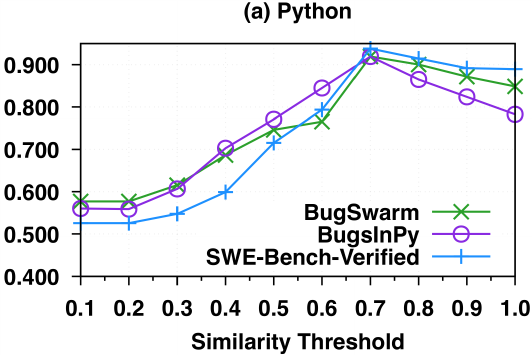}}
    \subfigure{\includegraphics[width=0.49\linewidth]{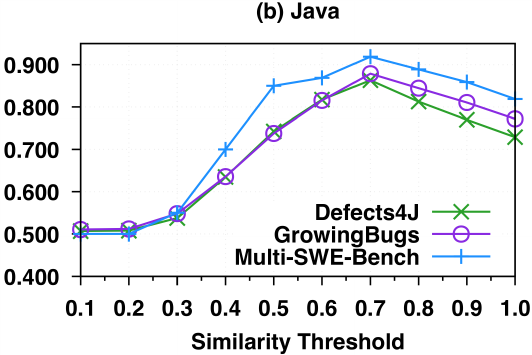}}
    \caption{AUC under different similarity thresholds.}
    \label{fig:sim}
\end{figure}

\noindent\textbf{Similarity Threshold}. In the patch presence test, the similarity threshold governs the matching process between the patch and the target snapshot code. Analogous to the default hit probability, setting this parameter too low makes the detector overly lenient and increases the false positive rate. Figure~\ref{fig:sim} depicts the impact of this threshold on patch detection performance. The results indicate consistent trends for both Python and Java, where optimal performance is achieved at a threshold of 0.7, below which detection performance degrades rapidly.

\begin{figure}
    \centering
    \revise{\includegraphics[width=0.98\linewidth]{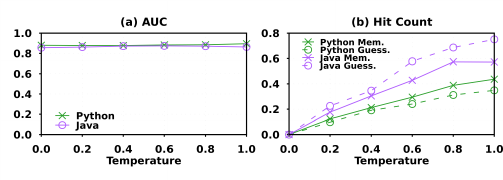}}
    \caption{\revise{AUC and relative hit-count changes across decoding temperatures.}}
    \label{fig:decoding_sensitivity}
\end{figure}

\noindent\textbf{\revise{Decoding Sensitivity.}} 
\revise{We study how decoding settings affect Stage I, where exact target reproduction is used to count hits. We focus on decoding temperature, which directly controls sampling randomness in our setting. A higher temperature generally increases output diversity and may therefore affect both memorization-driven hits and guessing-driven hits. On BugSwarm (Python) and GrowingBugs (Java), however, varying the temperature changes the AUC by at most 0.02 for both languages, as shown in Figure~\ref{fig:decoding_sensitivity}. The hit-count trends suggest that higher temperatures increase memorization-driven hits on included repositories and guessing-driven hits on excluded repositories at similar magnitudes, causing their effects to cancel out. Therefore, we set the default temperature to 1.0 for all experiments.}

\section{Empirical Study}
\label{sec:empirical_study}

We leverage \TN{} to empirically investigate the impact of data contamination on LLM bug-fixing evaluation. We \revise{select} SWE-bench Verified\revise{~\cite{chowdhury2024swebenchverified}} and the Java subset of Multi-SWE-bench\revise{~\cite{zan2025multi}} as subjects, given their transparent official leaderboards\revise{~\cite{swebench_leaderboards, multiswebench_leaderboard}} and comprehensive submission logs\revise{~\cite{swebench_experiments, multiswebench_experiments}}. Both benchmarks categorize task difficulty, ranging from \texttt{\textless{} 15 min} to \texttt{\textgreater{} 4 h} in SWE-bench Verified, and \texttt{Easy} to \texttt{Hard} in Multi-SWE-bench. Our analysis encompasses 159 submissions (123 for SWE-bench Verified and 36 for Multi-SWE-bench) that disclose their underlying models as of October 2025. For models without published training cutoff dates, we assume their cutoff dates are three months before the model release date. Notably, while most benchmark patches predate 2023, the training cutoff dates for nearly all SOTA models except GPT-4 fall in 2024 or later. \revise{For this empirical study, the default hit probability $P_0$ is calibrated separately for each target model using post-cutoff repositories, while the similarity threshold is obtained from The Stack v2 and reused because Stage~II relies on model-independent code similarity rather than model outputs.}

\revise{In our results, Stage~I detects the repositories of both benchmarks as present in the training set of every analyzed closed-source model at the $p < 0.05$ threshold, which is expected because all these repositories are popular open-source libraries. Nevertheless, Stage~I remains necessary: it provides model-specific repository-presence evidence rather than assuming inclusion from repository popularity alone, and this evidence is the prerequisite for applying Stage~II. In particular, closed-source training corpora are not publicly verifiable and may exclude even popular repositories due to licensing or other curation policies~\cite{deng2025swebenchpro}.}

Overall, closed-source models exhibit a significant performance drop on decontaminated samples. This confirms that even for benchmarks where most samples predate the cutoff, our method effectively identifies high-probability contaminated samples while retaining the majority of potentially clean ones, thus facilitating a credible evaluation of closed-source models.

\begin{table*}[t]
\centering
\setlength{\tabcolsep}{2pt}
\renewcommand{\arraystretch}{1.15}
\caption{\revise{Resolution rates} of top-10 submissions on \revise{the} SWE-bench Verified leaderboard (before vs. after applying \TN{}).}
\label{tab:top10-swe-bench-verified}
\footnotesize
\begin{tabular*}{\textwidth}{@{\extracolsep{\fill}} l l l l l l l @{}}
\toprule
\multicolumn{1}{c}{\textbf{Framework}} &
\multicolumn{1}{c}{\textbf{Model}} &
\multicolumn{1}{c}{\textbf{Overall}} &
\multicolumn{1}{c}{\textbf{\textless{} 15 min}} &
\multicolumn{1}{c}{\textbf{15 -- 60 min}} &
\multicolumn{1}{c}{\textbf{1 -- 4 h}} &
\multicolumn{1}{c}{\textbf{\textgreater{} 4 h}} \\
\midrule
TRAE        & Doubao-Seed        & \BAdown{394/500}{130/174}{5.2\%}   & \BAdown{169/194}{48/57}{3.3\%}   & \BAup{205/261}{73/91}{2.1\%}    & \BAdown{19/42}{9/25}{20.4\%}   & \BAdown{1/3}{0/1}{100.0\%} \\
EPAM        & Claude-4-Sonnet    & \BAdown{384/500}{122/174}{8.7\%}   & \BAdown{175/194}{50/57}{2.8\%}   & \BAdown{194/261}{66/91}{2.4\%}  & \BAdown{14/42}{6/25}{28.0\%}   & \BAdown{1/3}{0/1}{100.0\%} \\
Lingxi      & Claude-4-Sonnet    & \BAdown{373/500}{120/174}{7.6\%}   & \BAdown{170/194}{49/57}{1.9\%}   & \BAdown{186/261}{64/91}{1.3\%}  & \BAdown{16/42}{7/25}{26.5\%}   & \BAdown{1/3}{0/1}{100.0\%} \\
Refact      & Claude-4-Sonnet    & \BAdown{372/500}{120/174}{7.3\%}   & \BAup{169/194}{50/57}{0.7\%}     & \BAdown{186/261}{63/91}{2.9\%}  & \BAdown{16/42}{7/25}{26.5\%}   & \BAdown{1/3}{0/1}{100.0\%} \\
Tools       & Claude-4-Opus      & \BAdown{366/500}{116/174}{8.9\%}   & \BAdown{163/194}{46/57}{4.0\%}   & \BAdown{184/261}{62/91}{3.4\%}  & \BAdown{18/42}{8/25}{25.3\%}   & \BAdown{1/3}{0/1}{100.0\%} \\
Tools       & Claude-4-Sonnet    & \BAdown{362/500}{114/174}{9.5\%}   & \BAdown{160/194}{42/57}{10.7\%}  & \BAdown{187/261}{65/91}{0.3\%}  & \BAdown{15/42}{7/25}{21.6\%}   & \BAdown{0/3}{0/1}{0.0\%} \\
OpenHands   & GPT-5              & \BAdown{359/500}{96/149}{10.3\%}   & \BAdown{165/194}{38/48}{6.9\%}   & \BAdown{179/261}{53/79}{2.2\%}  & \BAdown{14/42}{5/21}{28.6\%}   & \BAdown{1/3}{0/1}{100.0\%} \\
Prometheus  & GPT-5              & \BAdown{356/500}{96/149}{9.5\%}    & \BAdown{159/194}{34/48}{13.6\%}  & \BAup{182/261}{56/79}{1.7\%}     & \BAdown{15/42}{6/21}{20.0\%}   & \BAdown{0/3}{0/1}{0.0\%} \\
Moatless    & Claude-4-Sonnet    & \BAdown{354/500}{112/174}{9.1\%}   & \BAdown{163/194}{45/57}{6.0\%}   & \BAdown{177/261}{61/91}{1.2\%}  & \BAdown{14/42}{6/25}{28.0\%}   & \BAdown{0/3}{0/1}{0.0\%} \\
OpenHands   & Claude-4-Sonnet    & \BAdown{352/500}{108/174}{11.8\%}  & \BAdown{164/194}{43/57}{10.8\%}  & \BAdown{173/261}{59/91}{2.2\%}  & \BAdown{14/42}{6/25}{28.0\%}   & \BAdown{1/3}{0/1}{100.0\%} \\
\bottomrule
\end{tabular*}
\end{table*}

\begin{table*}[t]
\centering
\setlength{\tabcolsep}{2pt}
\renewcommand{\arraystretch}{1.15}
\caption{\revise{Resolution rates} of top-10 submissions on \revise{the} Multi-SWE-bench leaderboard (before vs. after applying \TN{}).}
\label{tab:multi-swe-top10}
\footnotesize
\begin{tabular*}{\textwidth}{@{\extracolsep{\fill}} l l l l l l @{}}
\toprule
\multicolumn{1}{c}{\textbf{Framework}} &
\multicolumn{1}{c}{\textbf{Model}} &
\multicolumn{1}{c}{\textbf{Overall}} &
\multicolumn{1}{c}{\textbf{Easy}} &
\multicolumn{1}{c}{\textbf{Medium}} &
\multicolumn{1}{c}{\textbf{Hard}} \\
\midrule
SWE-agent  & Gemini-2.5-Pro    & \BAdown{37/128}{12/58}{28.4\%} & \BAdown{16/27}{7/15}{21.2\%} & \BAdown{21/65}{5/23}{32.7\%} & \BAdown{0/36}{0/20}{0.0\%} \\
SWE-agent  & Claude-3.7        & \BAdown{30/128}{11/58}{19.1\%} & \BAdown{12/27}{6/15}{10.0\%} & \BAdown{18/65}{5/23}{21.5\%} & \BAdown{0/36}{0/20}{0.0\%} \\
Agentless  & DeepSeek\mbox{-}R1& \BAdown{29/128}{9/58}{31.5\%}  & \BAdown{14/27}{6/15}{22.9\%} & \BAdown{15/65}{3/23}{43.5\%} & \BAdown{0/36}{0/20}{0.0\%} \\
SWE-agent  & OpenAI o1                & \BAdown{28/128}{14/71}{9.9\%}  & \BAdown{13/27}{8/18}{7.7\%}  & \BAdown{15/65}{6/31}{16.1\%} & \BAdown{0/36}{0/22}{0.0\%} \\
OpenHands  & Claude-3.7        & \BAdown{28/128}{10/58}{21.2\%} & \BAdown{13/27}{6/15}{16.9\%} & \BAdown{15/65}{4/23}{24.6\%} & \BAdown{0/36}{0/20}{0.0\%} \\
Agentless  & Gemini-2.5-Pro    & \BAdown{28/128}{9/58}{29.1\%}  & \BAdown{14/27}{7/15}{10.0\%} & \BAdown{14/65}{2/23}{59.6\%} & \BAdown{0/36}{0/20}{0.0\%} \\
Agentless  & OpenAI o1                & \BAdown{27/128}{13/71}{13.2\%} & \BAdown{11/27}{7/18}{4.5\%}  & \BAdown{16/65}{6/31}{21.4\%} & \BAdown{0/36}{0/22}{0.0\%} \\
SWE-agent  & Claude-3.5        & \BAdown{26/128}{8/59}{33.2\%}  & \BAdown{13/27}{6/15}{16.9\%} & \BAdown{13/65}{2/24}{58.3\%} & \BAdown{0/36}{0/20}{0.0\%} \\
SWE-agent  & OpenAI o3               & \BAdown{21/128}{7/58}{26.4\%}  & \BAdown{9/27}{4/15}{20.0\%}  & \BAdown{12/65}{3/23}{29.3\%} & \BAdown{0/36}{0/20}{0.0\%} \\
SWE-agent  & Llama-4           & \BAdown{20/128}{4/58}{55.9\%}  & \BAdown{8/27}{3/15}{32.5\%}  & \BAdown{12/65}{1/23}{76.4\%} & \BAdown{0/36}{0/20}{0.0\%} \\
\bottomrule
\end{tabular*}
\end{table*}

\subsection{Resolution Rate Discrepancy} Table~\ref{tab:top10-swe-bench-verified} and Table~\ref{tab:multi-swe-top10} show the resolution rates for the \revise{top-10} submissions on \revise{the} SWE-bench Verified and Multi-SWE-bench leaderboards before and after excluding contaminated cases using \TN{}.

\textit{\textbf{Finding 1.} The impact of data contamination on model evaluation increases with the difficulty of bug fixing.} \revise{For SWE-bench Verified, the relative drops for all top-10 submissions on the \texttt{1-4 h} subset exceed 20\%, compared to relative drops below 10\% on both the \texttt{\textless{}15 min} and \texttt{15-60 min} subsets. For Multi-SWE-bench, the average relative drop of the top-10 submissions on the \texttt{Medium} subset is about 40\%, compared to 20\% on the \texttt{Easy} subset.} This suggests that \revise{models increasingly rely} on memorization as bug-solving difficulty increases, \revise{whereas success on simpler bugs indicates stronger generalization.}

\textit{\textbf{Finding 2.} The severity of the impact of data contamination varies across different benchmarks.} The overall resolution rates of the top-10 submissions drop by up to 55.9\% on Multi-SWE-bench, compared to a drop of up to 11.8\% on SWE-bench Verified. This indicates that data contamination has a more severe impact on evaluation in Multi-SWE-bench than in SWE-bench Verified, since Multi-SWE-bench samples are generally harder to resolve, and its submissions have not yet employed more advanced model families such as Claude-4~\cite{claude4} and GPT-5~\cite{gpt5}. 

\textit{\textbf{Finding 3.} The impact of data contamination is consistent across different agent-based frameworks using the same underlying model.} Among the top-10 submissions on the SWE-bench Verified leaderboard, 7 use Claude-4 family models, with their overall relative drop ranging from 8\% to 11\%. In contrast, among the top-10 submissions on Multi-SWE-bench, the maximum difference in relative drop across frameworks using the same underlying model is only 3.3\%.

\begin{table*}[t]
\centering
\setlength{\tabcolsep}{3pt}
\renewcommand{\arraystretch}{1.12}
\caption{Top-1 resolution rates of major LLMs on SWE-bench Verified by release date (before vs. after BenchChecker).}
\label{tab:model-resolve-rates-swe-bench-verified}
\footnotesize
\begin{tabularx}{\textwidth}{@{}>{\raggedright\arraybackslash}X l l l l l l@{}}
\toprule
\multicolumn{1}{c}{\textbf{Release Date}} & \multicolumn{1}{c}{\textbf{Model}} &
\multicolumn{1}{c}{\textbf{Overall}} & \multicolumn{1}{c}{\textbf{\textless{} 15 min}} &
\multicolumn{1}{c}{\textbf{15 -- 60 min}} & \multicolumn{1}{c}{\textbf{1 -- 4 h}} &
\multicolumn{1}{c}{\textbf{\textgreater{} 4 h}} \\
\midrule
2023-03 & GPT-4 & \BAdown{112/500}{17/132}{42.5\%} & \BAdown{71/194}{12/41}{20.0\%} & \BAdown{41/261}{5/72}{55.8\%} & \BAdown{0/42}{0/18}{0.0\%} & \BAdown{0/3}{0/1}{0.0\%} \\
2024-03 & Claude-3 & \BAdown{79/500}{12/128}{40.7\%} & \BAdown{53/194}{8/43}{31.9\%} & \BAdown{26/261}{4/65}{38.2\%} & \BAdown{0/42}{0/19}{0.0\%} & \BAdown{0/3}{0/1}{0.0\%} \\
2024-04 & Llama-3 & \BAdown{206/500}{36/132}{33.8\%} & \BAdown{111/194}{19/44}{24.5\%} & \BAdown{94/261}{17/68}{30.6\%} & \BAdown{1/42}{0/19}{100.0\%} & \BAdown{0/3}{0/1}{0.0\%} \\
2024-05 & GPT-4o & \BAdown{236/500}{43/129}{29.4\%} & \BAdown{130/194}{23/44}{22.0\%} & \BAdown{101/261}{19/65}{24.5\%} & \BAdown{5/42}{1/19}{55.8\%} & \BAdown{0/3}{0/1}{0.0\%} \\
2024-06 & Claude-3.5 & \BAdown{314/500}{73/141}{17.6\%} & \BAdown{158/194}{34/48}{13.0\%} & \BAdown{146/261}{36/71}{9.4\%} & \BAdown{9/42}{3/21}{33.3\%} & \BAdown{1/3}{0/1}{100.0\%} \\
2024-09 & OpenAI o1 & \BAdown{323/500}{71/129}{14.8\%} & \BAdown{150/194}{28/44}{17.7\%} & \BAdown{162/261}{40/65}{0.9\%} & \BAdown{10/42}{3/21}{33.7\%} & \BAdown{1/3}{0/1}{100.0\%} \\
2024-11 & Qwen2.5-Coder & \BAdown{164/500}{34/136}{23.8\%} & \BAdown{101/194}{21/45}{10.4\%} & \BAdown{61/261}{12/68}{24.5\%} & \BAdown{2/42}{1/21}{0.0\%} & \BAdown{0/3}{0/1}{0.0\%} \\
2024-12 & Gemini-2.0 & \BAdown{261/500}{59/149}{24.1\%} & \BAdown{135/194}{27/48}{19.2\%} & \BAdown{122/261}{32/79}{13.3\%} & \BAdown{4/42}{0/21}{100.0\%} & \BAdown{0/3}{0/1}{0.0\%} \\
2024-12 & DeepSeek-V3 & \BAdown{210/500}{38/144}{37.2\%} & \BAdown{116/194}{20/48}{30.3\%} & \BAdown{90/261}{17/74}{33.4\%} & \BAdown{4/42}{1/21}{50.0\%} & \BAdown{0/3}{0/1}{0.0\%} \\
2025-02 & Claude-3.7 & \BAdown{316/500}{83/154}{14.7\%} & \BAdown{154/194}{36/52}{12.8\%} & \BAdown{152/261}{44/80}{5.6\%} & \BAdown{9/42}{3/21}{33.3\%} & \BAdown{1/3}{0/1}{100.0\%} \\
2025-04 & OpenAI o3 & \BAdown{341/500}{83/141}{13.7\%} & \BAdown{158/194}{34/48}{13.0\%} & \BAdown{171/261}{45/71}{3.3\%} & \BAdown{11/42}{4/21}{27.3\%} & \BAdown{1/3}{0/1}{100.0\%} \\
2025-05 & Doubao-Seed & \BAdown{394/500}{130/174}{5.2\%} & \BAdown{169/194}{48/57}{3.3\%} & \BAup{205/261}{73/91}{2.1\%} & \BAdown{19/42}{9/25}{20.4\%} & \BAdown{1/3}{0/1}{100.0\%} \\
2025-05 & Claude-4 & \BAdown{384/500}{122/174}{8.7\%} & \BAdown{175/194}{50/57}{2.8\%} & \BAdown{194/261}{66/91}{2.4\%} & \BAdown{14/42}{6/25}{28.0\%} & \BAdown{1/3}{0/1}{100.0\%} \\
2025-07 & Qwen3-Coder & \BAdown{348/500}{109/174}{10.0\%} & \BAdown{165/194}{46/57}{5.1\%} & \BAdown{168/261}{57/91}{2.7\%} & \BAdown{14/42}{6/25}{28.0\%} & \BAdown{1/3}{0/1}{100.0\%} \\
2025-08 & GPT-5 & \BAdown{359/500}{96/149}{10.3\%} & \BAdown{165/194}{38/48}{6.9\%} & \BAdown{179/261}{53/79}{2.2\%} & \BAdown{14/42}{5/21}{28.6\%} & \BAdown{1/3}{0/1}{100.0\%} \\
\bottomrule
\end{tabularx}
\end{table*}

\begin{table*}[t]
\centering
\setlength{\tabcolsep}{3pt}
\renewcommand{\arraystretch}{1.12}
\caption{Top-1 resolution rates of major LLMs on Multi-SWE-bench by release date (before vs.\ after BenchChecker).}
\label{tab:multi-swe-model-resolve-rates}
\footnotesize
\begin{tabular*}{\textwidth}{@{\extracolsep{\fill}} l l l l l l @{}}
\toprule
\multicolumn{1}{c}{\textbf{Release Date}} &
\multicolumn{1}{c}{\textbf{Model}} &
\multicolumn{1}{c}{\textbf{Overall}} &
\multicolumn{1}{c}{\textbf{Easy}} &
\multicolumn{1}{c}{\textbf{Medium}} &
\multicolumn{1}{c}{\textbf{Hard}} \\
\midrule
2024-05 & GPT-4o          & \BAdown{16/128}{5/71}{43.7\%}   & \BAdown{6/27}{3/18}{25.0\%}    & \BAdown{10/65}{2/31}{58.1\%}   & \BAdown{0/36}{0/22}{0.0\%} \\
2024-06 & Claude-3.5      & \BAdown{26/128}{8/59}{33.2\%}   & \BAdown{13/27}{6/15}{16.9\%}   & \BAdown{13/65}{2/24}{58.3\%}   & \BAdown{0/36}{0/20}{0.0\%} \\
2024-09 & OpenAI o1       & \BAdown{28/128}{14/71}{9.9\%}   & \BAdown{13/27}{8/18}{7.7\%}    & \BAdown{15/65}{6/31}{16.1\%}   & \BAdown{0/36}{0/22}{0.0\%} \\
2024-11 & Qwen2.5-Coder   & \BAdown{14/128}{5/58}{21.2\%}   & \BAdown{9/27}{4/15}{20.0\%}    & \BAdown{5/65}{1/23}{43.5\%}    & \BAdown{0/36}{0/20}{0.0\%} \\
2024-12 & DeepSeek-V3     & \BAdown{15/128}{4/58}{41.1\%}   & \BAdown{9/27}{3/15}{40.0\%}    & \BAdown{6/65}{1/23}{52.9\%}    & \BAdown{0/36}{0/20}{0.0\%} \\
2025-01 & DeepSeek-R1     & \BAdown{29/128}{9/58}{31.5\%}   & \BAdown{14/27}{6/15}{22.9\%}   & \BAdown{15/65}{3/23}{43.5\%}   & \BAdown{0/36}{0/20}{0.0\%} \\
2025-01 & Doubao-1.5      & \BAdown{17/128}{6/58}{22.1\%}   & \BAdown{9/27}{4/15}{20.0\%}    & \BAdown{8/65}{2/23}{29.3\%}    & \BAdown{0/36}{0/20}{0.0\%} \\
2025-01 & OpenAI o3-mini  & \BAdown{21/128}{7/58}{26.4\%}   & \BAdown{9/27}{4/15}{20.0\%}    & \BAdown{12/65}{3/23}{29.3\%}   & \BAdown{0/36}{0/20}{0.0\%} \\
2025-02 & Claude-3.7      & \BAdown{30/128}{11/58}{19.1\%}  & \BAdown{12/27}{6/15}{10.0\%}   & \BAdown{18/65}{5/23}{21.5\%}   & \BAdown{0/36}{0/20}{0.0\%} \\
2025-03 & Gemini-2.5-Pro  & \BAdown{37/128}{12/58}{28.4\%}  & \BAdown{16/27}{7/15}{21.2\%}   & \BAdown{21/65}{5/23}{32.7\%}   & \BAdown{0/36}{0/20}{0.0\%} \\
2025-04 & Llama-4         & \BAdown{20/128}{4/58}{55.9\%}   & \BAdown{8/27}{3/15}{32.5\%}    & \BAdown{12/65}{1/23}{76.4\%}   & \BAdown{0/36}{0/20}{0.0\%} \\
\bottomrule
\end{tabular*}
\end{table*}

\vspace{-0.35em}
\subsection{Resolution Rate Discrepancy in Timeline}

Tables~\ref{tab:model-resolve-rates-swe-bench-verified} and~\ref{tab:multi-swe-model-resolve-rates} present changes in resolution rates for the best submission associated with each model on the SWE-bench Verified and Multi-SWE-bench leaderboards, sorted by the models' release dates.

\textit{\textbf{Finding 4.} The impact of data contamination on model evaluation decreases \revise{as models become more recent}.} The SWE-bench Verified \revise{leaderboard} displays a clear trend: the overall discrepancy in resolution rate decreases from 42.5\% in 2023 to 5.2\% in 2025. Specifically, in its \texttt{15-60 min} subset, the \revise{discrepancy} decreases from 55.8\% to nearly 0\%. This indicates a positive signal: model generalization gradually improves over time, particularly for bugs with lower difficulty. \revise{A similar trend is observed on Multi-SWE-bench, although the trend is less pronounced due to its relatively smaller number of submissions.} 

\textit{\textbf{Finding 5.} Models with stronger reported performance show stronger generalization.} By comparing the \revise{reported} performance of these models on the SWE-bench Verified and Multi-SWE-bench leaderboards, models with higher reported resolution rates exhibit smaller drops in resolution rates after contamination removal. This indicates that the model's generalization capability is strongly correlated with its reported performance. Therefore, despite the issue of data contamination, the benchmarks can still reflect genuine trends in model performance improvement.

\vspace{-0.35em}
\subsection{Bug-Fixing Patterns of SOTA LLMs}
\label{sec:bug_patterns}

We analyze the capability limits of SOTA models in bug-fixing tasks and identify patterns of bug samples that SOTA models succeed or fail to fix. Specifically, we focus on three categories of samples: (1) seen patches that models fail to generate, (2) unseen patches that models \revise{succeed in generating}, and (3) unseen patches that models fail to generate. For each category, we analyze samples that consistently fall into that category across the top-10 submissions on the SWE-bench Verified leaderboard.

\begin{figure}[t]
\begin{lstlisting}[style=arxivdiff]
// Gold patch:
--- a/django/utils/timezone.py
+++ b/django/utils/timezone.py
@@ -72,8 +72,11 @@ def get_current_timezone_name():
 
 def _get_timezone_name(timezone):
+    return timezone.tzname(None) or 
+           str(timezone)

// Model patch:
--- a/django/db/backends/postgresql/operations.py
+++ b/django/db/backends/postgresql/operations.py
@@ -44,10 +44,12 @@ class DatabaseOperations(BaseDatabaseOperations):
 
 def _prepare_tzname_delta(self, tzname):
-   if '+' in tzname:
-       return tzname.replace('+', '-')
-   elif '-' in tzname:
-       return tzname.replace('-', '+')
+   if tzname and len(tzname) > 0 and tzname[0] in '+-':
+       if all(c in '0123456789:' for c in tzname[1:]):
+       return ('-' if tzname[0] == '+' else '+') + tzname[1:]
    return tzname
 
\end{lstlisting}
\caption{\revise{Gold patch and model-generated patch (TRAE/Doubao-Seed-Code~\cite{seed2025seed}) for the instance \texttt{django\_\_django-14792} on SWE-bench Verified.}}
\label{lst:django__django-14792}
\vspace{-1.0em}
\end{figure}

\textit{\textbf{Finding 6.} Bug-localization failure is the most common reason models fail to generate seen patches.} We find that the most common failure pattern when models fail on bugs whose patches are seen is buggy code localization failure (16/30). This occurs because issue descriptions typically do not explicitly indicate buggy-code locations, so models must infer them. When the model's buggy code localization is inaccurate, the memorized patch is not applied to the correct code location. Figure~\ref{lst:django__django-14792} presents an example labeled as \texttt{\textless{} 15 min} difficulty: the gold patch requires only a \revise{one-line} fix to the \texttt{\_get\_timezone\_name} function; however, the model incorrectly localizes the bug to the \texttt{\_prepare\_tzname\_delta} function, resulting in a failed repair. Another common failure pattern is partial bug fixes (8/30), which often stem from the model's inability to localize all erroneous code locations.

 \textit{\textbf{Finding 7.} \revise{Bug-triggering cases and solution hints in issue descriptions make bugs easier for models to fix.}} Analysis of samples successfully resolved by \revise{models} reveals that the majority ($49/61$) involve single-file modifications, while over half ($35/61$) contain specific bug-triggering scenarios. These findings indicate that models are most effective when the repair scope is constrained and diagnostic information is clearly articulated. Notably, about one-third of the samples ($21/61$) directly provide solution hints, substantially lowering the complexity of the task. Figure~\ref{lst:pallets__flask-5014} illustrates a representative case in which the issue description explicitly suggests using \texttt{raise ValueError}, a recommendation that aligns exactly with the implemented patch.

\textit{\textbf{Finding 8.} Models still face challenges in localizing and fixing bugs across multiple files.} For samples with unseen patches that models fail to repair, we observe that none of these samples provides explicit solution hints in their issue descriptions (25/25), and the majority require cross-file modifications (15/25). This suggests that while current LLMs can effectively handle self-contained bugs within a single file, they struggle to capture dependencies and interactions that span multiple files or modules. \revise{Such cases often demand capabilities that remain challenging for today's models, including a deeper understanding of the project's architecture, build dependencies, and execution flow.}

\section{Discussion}

\noindent \textbf{Transferability to Other Tasks.} We extended \TN{} beyond bug fixing to other code-related tasks by applying it to the code-completion benchmark RepoBench~\cite{liu2023repobench,repobench2024}. Our analysis indicates that the inclusion of a sample's repository in the training set increases the exact match rate by up to 60\% compared to unseen repositories. This finding confirms that repository presence is a robust indicator of potential contamination and highlights the critical role of the first-stage repository presence test in \TN{} (see Appendix~\ref{sec:appendix_b} for details). 

\noindent \textbf{Transferability to Other Data Modalities.} Repetitiveness can be adapted through co-occurrence patterns~\cite{harris1957co}; for instance, in book datasets~\cite{chang2023speak}, frequently co-occurring character names within paragraphs form identifiable co-occurrence patterns. Distinctiveness similarly applies across modalities, as unique character names or an artist’s stylistic signatures can serve as distinctive features. Context-independence may play a smaller role outside code, since programming languages are inherently structured~\cite{lau2006selecting}, providing stronger contextual cues than natural language or artwork, where the relationship between context and target is generally weaker.

\begin{figure}[t]
\begin{lstlisting}[style=arxivdiff]
// Title: Require a non-empty name for Blueprints
// Description: Things do not work correctly if a 
// Blueprint is given an empty name (e.g. #4944).
// It would be helpful if a `ValueError` was raised 
// when trying to do that.
--- a/src/flask/blueprints.py
+++ b/src/flask/blueprints.py
@@ -190,6 +190,9 @@ def __init__(
    root_path=root_path,)
 
+   if not name:
+       raise ValueError("'name' may not be empty.")
+
    if "." in name:
        raise ValueError("'name' may not contain a dot '.' character.")
\end{lstlisting}
\caption{\revise{Issue description and patch for the instance \texttt{pallets\_\_flask-5014} on SWE-bench Verified.}}
\label{lst:pallets__flask-5014}
\end{figure}

\noindent \textbf{Limitations and Future Work.}
\revise{
Our method has three main limitations. First, for scalability, Stage~I assesses target distinctiveness primarily at the lexical rather than structural level. Second, Stage~II assumes that a contaminated patch remains observable in the cutoff snapshot. If a patch appears in earlier training snapshots but is removed or substantially refactored before the cutoff snapshot, \TN{} may underestimate contamination. Our dataset analysis suggests that such cases account for only a small fraction of samples, as discussed in Section~\ref{sec:datasets}. Third, \TN{} does not directly cover contamination through artifacts without comparable code, such as issue descriptions or natural-language discussions. We leave these cases to future work. Notably, for training strategies that intentionally include full repository commit histories~\cite{yang2025swe}, a conservative mitigation is to apply the Stage~I repository presence test of \TN{} and exclude samples from repositories detected as present in the training corpus.
}

\section{Related Work}

\noindent \textbf{Data Contamination Study in Bug-Fixing Benchmarks.} Recent research has begun to study data contamination in bug-fixing benchmarks. Ramos et al.~\cite{ramos2024large} used metrics such as negative log-likelihood to assess LLM memorization on Defects4J~\cite{just2014defects4j}. Liang et al.~\cite{liang2025swe} observed that several SOTA models can achieve high file-path identification accuracy on SWE-bench Verified without repository-structure information. \revise{Kong et al.~\cite{kong2025demystifying} assessed whether the model could still produce the original patch after the code had been heavily mutated in Defects4J. Recent cybersecurity benchmarks have also reported limited cutoff effects on vulnerability detection/patching and PoC-generation tasks~\cite{zhang2025bountybench,wang2026cybergym}.}

\noindent \textbf{Patch Presence Test.} The patch presence test usually refers to the technique used to determine whether a known security patch has been applied to an executable or firmware, without relying on version numbers or changelogs~\cite{zhang2018precise,he2024bingo}. It focuses on detecting concrete evidence of the patch itself, rather than inferring it from software versions. This approach enables independent and measurable security assessments, particularly in scenarios involving closed-source components, third-party libraries, and complex supply chains~\cite{dai2020bscout}. It is widely used in mobile and IoT firmware~\cite{sun2021osprey}, the Android/Java ecosystem~\cite{dai2020bscout}, C/C++ system software~\cite{zhang2018precise}, and in responding to supply chain risks and identifying one-day vulnerabilities~\cite{zhang2021investigation}.

\section{Conclusion}

We introduced \TN{}, a forensics-based two-stage framework that detects data contamination in bug-fixing benchmarks by combining repository-level and patch-level verification while \revise{relying only on textual outputs}. Our evaluation demonstrates that \TN{} consistently outperforms existing baselines and achieves a robust AUC of approximately 0.9 across multiple benchmarks. Our analysis further reveals that excluding contaminated samples leads to a drop of over 20\% in resolution rates for medium-difficulty tasks, which indicates significant overestimation in current LLM evaluations.

\section*{Ethical Considerations}

\revise{This paper evaluates data contamination in bug-fixing benchmarks for large language models (LLMs). Specifically, we describe the impacts on three stakeholder groups along with our mitigation strategies.}

\revise{\noindent \textbf{Benchmark Creators and Maintainers.} This group creates benchmark datasets and maintains leaderboards for evaluating the performance of LLMs. Our findings might temporarily raise doubts about the accuracy of their current leaderboard rankings. To mitigate this, maintainers can use our tool to automatically filter out contaminated samples. This helps strengthen the credibility of their leaderboards. We will contact the maintainers of the SWE-bench and Multi-SWE-bench leaderboards to notify them about our study results.}

\revise{\noindent \textbf{LLM Vendors and Developers.} LLM vendors and developers are the individuals and organizations responsible for building, training, and releasing LLMs. Our study on data contamination might lower their models' reported rankings and performance on public leaderboards. However, our evaluation remains neutral and does not target specific LLMs. Our tool helps reveal the real capabilities of LLMs in bug-fixing tasks.}

\revise{\noindent \textbf{End Users and Enterprises.} End users may rely on leaderboards to choose AI coding tools. Learning that scores are inflated might cause a temporary loss of trust. However, we believe sharing these results transparently is the best mitigation. Hiding these facts could lead users to over-rely on inflated benchmark scores in production settings.}

\revise{%
We use benchmark datasets, repositories, and leaderboard submissions that are publicly released for software development, are widely used in the community, and contain no sensitive data. The main risk of conducting this study is that an incorrect contamination label may unfairly affect a benchmark or model submission. We mitigate this risk by using transparent criteria in our open-source implementation, conducting error analysis, and engaging with benchmark maintainers to review the findings and potential mitigations.}

\revise{%
Given the low residual risk after these safeguards and the clear benefit of improving benchmark credibility, we decided to conduct and publish this study. Furthermore, it offers a useful tool to help maintainers strengthen benchmark credibility and suggests a better way for the community to conduct repository-level benchmarks while reducing the effect of data contamination.
}

\section*{Open Science}

The code and evaluation data are available at \url{https://zenodo.org/records/20385623}.
\section*{Acknowledgments}
This work is partially supported by the U.S. Office of Naval Research under grant N00014-23-1-2122, CCI AI for Cybersecurity N-3Q25-002, and a gift from VISA Inc.

\bibliographystyle{plainurl}
\bibliography{Reference}
\appendix
\setcounter{section}{0}
\section{Case Study on RepoBench}
\label{sec:appendix_b}

\revise{We also apply \TN{} to RepoBench~\cite{liu2023repobench} to study how sample repository inclusion affects repository-level code generation. RepoBench is a widely used repository-level benchmark for cross-file retrieval and generation with dynamic updates designed to reduce contamination risk. We evaluate CodeLlama-34B~\cite{roziere2023code}, GPT-4o~\cite{achiam2023gpt}, and DeepSeekCoderV2-Lite~\cite{zhu2024deepseek} on the latest RepoBench v1.1 Cross-File-First split~\cite{repobench2024}. This split contains repositories created between October 6 and December 31, 2023, and has minimal overlap with the models' reported cutoffs. The task asks a model to predict the first cross-file line that appears before any local reference to the module or symbol, testing cross-file dependency handling without local cues. We use 580 repositories and 5,054 questions.}

\begin{center}
    \centering
    \includegraphics[width=0.7\linewidth]{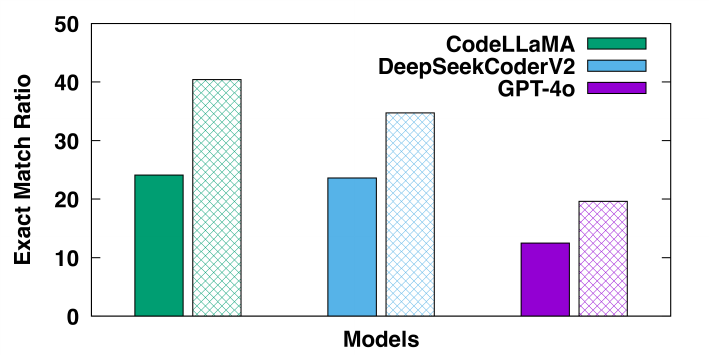}
    \captionof{figure}{\revise{Exact-match ratios on RepoBench repositories detected as included or excluded.}}
    \label{fig:cs2_em}
\end{center}

\begin{center}
\centering
\scriptsize
\captionof{table}{\revise{Number of Repositories Across Different $p$-Values}}
\label{tab:bench_contamination_detection}
\setlength{\tabcolsep}{5pt}
\begin{tabular}{lcccc}
\hline
\textbf{Metric} & \textbf{CodeLlama-34B} & \textbf{DeepSeekCoderV2-Lite} & \textbf{GPT-4o} \\
\hline
$p < 0.05$ & 78 & 82 & 68 \\
$p = 1.0$ & 292 & 278 & 290 \\
\hline
\end{tabular}
\end{center}

\revise{Table~\ref{tab:bench_contamination_detection} shows the number of repositories that \TN{} detects as high risk for inclusion ($p<0.05$). Across all three models, approximately 15\% of RepoBench repositories show strong signs of memorization despite being created after the models' reported cutoffs. This indicates that post-cutoff benchmark construction reduces but does not eliminate contamination risk, because new repositories can reuse older code, structure, or documentation. For example, \textit{wangxianhui111/xuechengzaixian}~\cite{xuechengzaixian2024} is detected for all three models; it aggregates several Java teaching projects, and a project with the same name and similar content has been available on GitHub since 2018~\cite{xuechengzaixianGaojie2024}.}

\section{\revise{Pseudocode of Core Algorithms}}
\label{sec:appendix_a}

\revise{Algorithms~\ref{alg:simhash_clustering}--\ref{alg:patch_block_locator} summarize \TN{}'s four core procedures: text-duplication clustering, word co-occurrence extraction, entropy scoring, and patch-code localization.}

\revise{
    \small
    \algrenewcommand\algorithmicindent{0.65em}
    \begin{enumerate}
        \setlength{\itemsep}{1pt}
        \setlength{\parsep}{0pt}
        \setlength{\topsep}{1pt}
        \setlength{\partopsep}{0pt}
        \item \textbf{SimHash Text-Duplication Detection.}\label{alg:simhash_clustering}
        \begin{algorithmic}
            \Statex \textbf{In:} texts $T$, threshold $\beta$; \textbf{Out:} clusters $C$
            \State $C \gets \{\}$
            \ForAll{$t \in T$}
                \State $v \gets \text{\_simhash}(t)$; $matched \gets \text{False}$
                \ForAll{$h \in \text{keys}(C)$}
                    \If{$\text{\_hamming\_distance}(v,h) \leq \beta$}
                        \State $C[h].\text{append}(t)$; $matched \gets \text{True}$; \textbf{break}
                    \EndIf
                \EndFor
                \If{not $matched$} \State $C[v] \gets [t]$ \EndIf
            \EndFor
            \State \Return $\text{values}(C)$
        \end{algorithmic}

        \item \textbf{PrefixSpan Word Co-occurrence Detection.}\label{alg:pattern_extraction}
        \begin{algorithmic}
            \Statex \textbf{In:} files $F$, words $W$; \textbf{Out:} patterns $P$
            \State $S \gets [\ ]$
            \ForAll{$f \in F$}
                \State $M \gets \text{\_nonzero\_frequencies}(W,f)$
                \State $S.\text{append}(\text{\_expand\_by\_frequency}(M))$
            \EndFor
            \State $P \gets \text{\_merge\_patterns}(\text{\_prefixspan}(S))$
            \State \Return $P$
        \end{algorithmic}

        \item \textbf{Target-Query Entropy Calculation.}\label{alg:loss_based_filtering}
        \begin{algorithmic}
            \Statex \textbf{In:} target $T$, query $Q$, model $M_r$; \textbf{Out:} entropy $H$
            \State $T_s \gets \text{\_sub\_word\_split}(T)$; $Q_s \gets \text{\_sub\_word\_split}(Q)$
            \State $queue \gets [Q]$; $L \gets [\ ]$
            \ForAll{$t \in T_s$}
                \If{$\exists q\in Q_s: t=q \lor \text{\_lexical\_match}(t,q)$}
                    \State $queue.\text{push}(\text{\_concat}(queue.\text{pop}(),t))$
                \Else
                    \State $L.\text{append}(\text{\_compute\_loss}(M_r, queue.\text{pop}(), t))$
                \EndIf
            \EndFor
            \State \Return $\max(L)$
        \end{algorithmic}

        \item \textbf{Patch Code-Block Localization.}\label{alg:patch_block_locator}
        \begin{algorithmic}
            \Statex \textbf{In:} patch $D$, repository $R$; \textbf{Out:} locations $L$
            \State $L \gets [\ ]$
            \ForAll{$hunk \in \text{\_parse\_patch}(D)$}
                \State $A \gets \text{\_extract\_added\_lines}(hunk)$; $f \gets hunk.\text{target\_file}$
                \State $P_a \gets \text{\_parse}(\text{\_join}(A))$; $R_a \gets \text{\_parse}(\text{\_read\_file}(R,f))$
                \ForAll{$\mathit{func} \in \text{\_find\_candidate\_functions}(P_a,R_a)$}
                    \State $F_a \gets \text{\_extract\_function\_ast}(R_a, \mathit{func})$
                    \State $m \gets \text{\_gumtree\_subtree\_match}(P_a,F_a)$
                    \If{$m.\text{similarity}>\theta$}
                        \State $L.\text{append}(\text{\_Location}(f, \mathit{func}, m))$
                        \State \textbf{break}
                    \EndIf
                \EndFor
            \EndFor
            \State \Return $L$
        \end{algorithmic}
    \end{enumerate}
}

\end{document}